\documentclass[prb,twocolumn,showpacs,preprintnumbers,amsmath,amssymb, superscriptaddress]{revtex4-2}   
\usepackage[utf8]{inputenc}
\usepackage{amsmath}    
\usepackage{amssymb}
\usepackage{graphicx}   
\usepackage{color}
\usepackage{xcolor}      
\usepackage{hyperref}   
\usepackage[normalem]{ulem}
\usepackage{bm}
\usepackage{fixmath}
\usepackage{dsfont}

\hypersetup{colorlinks,linkcolor=blue,urlcolor=blue,citecolor=blue}

\newcommand{\br}{{\bf{r} }}

\newcommand{\SUN}{{\textrm{SU}(N)}} 
\newcommand{\be}{\begin{equation}}
\newcommand{\ee}{\end{equation}}

\newcommand{\mysection}[1]{  {\null  \vskip0.2cm {\parindent=0pt \textbf{\large #1}}} \\}
\newcommand{\mysubsection}[1]{ {\null  \vskip0.1cm \parindent=0pt \textbf{#1}}}

\begin{document}
\title{Spectroscopic evidence for engineered hadron formation in repulsive \\ fermionic $\textrm{SU}(N)$ Hubbard Models}

\author{Mikl\'os Antal Werner}
\affiliation{MTA-BME Quantum Dynamics and Correlations Research Group, 
Department of Theoretical Physics, Institute of Physics, Budapest University of Technology and Economics, Műegyetem rkp. 3., H-1111 Budapest, Hungary}
\affiliation{MTA-BME Exotic Quantum Phases 'Lend\"ulet' Research Group, Department of Theoretical Physics, Institute of Physics, Budapest University of Technology and Economics, Műegyetem rkp. 3., H-1111 Budapest, Hungary}
\author{C\u at\u alin Pa\c scu Moca}
\affiliation{MTA-BME Quantum Dynamics and Correlations Research Group,
Department of Theoretical Physics, Institute of Physics, Budapest University of Technology and Economics, Műegyetem rkp. 3., H-1111 Budapest, Hungary}
\affiliation{Department of Physics, University of Oradea, 410087, Oradea, Romania}
\author{M\'arton Kormos}
\affiliation{MTA-BME Quantum Dynamics and Correlations Research Group,
Department of Theoretical Physics, Institute of Physics, Budapest University of Technology and Economics, Műegyetem rkp. 3., H-1111 Budapest, Hungary}
\author{\" Ors Legeza}
\affiliation{Strongly Correlated Systems 'Lend\" ulet' Research Group, Wigner Research Centre for Physics, P.O. Box 49, 1525 Budapest, Hungary}
\affiliation{Institute for Advanced Study, Technical University of Munich, Lichtenbergstrasse 2a, 85748 Garching, Germany}
 \author{Bal\'azs D\'ora}
\affiliation{MTA-BME Lend\"ulet Topology and Correlation Research Group,
Department of Theoretical Physics, Institute of Physics, Budapest University of Technology and Economics, Műegyetem rkp. 3., H-1111 Budapest, Hungary}
\author{Gergely Zar\'and}
\affiliation{MTA-BME Quantum Dynamics and Correlations Research Group,
Department of Theoretical Physics, Institute of Physics, Budapest University of Technology and Economics, Műegyetem rkp. 3., H-1111 Budapest, Hungary}
\affiliation{MTA-BME Exotic Quantum Phases 'Lend\"ulet' Research Group, Department of Theoretical Physics, Institute of Physics, Budapest University of Technology and Economics, Műegyetem rkp. 3., H-1111 Budapest, Hungary}
\date{\today}
\begin{abstract}

Particle formation represents a central theme in various branches of physics, often associated to confinement.
Here we show that dynamical hadron formation can be spectroscopically detected  in an ultracold atomic setting  
within the most paradigmatic and simplest model of condensed matter physics, the repulsive  $\SUN$ Hubbard model. 
By starting from an appropriately engineered  initial  state of the ${{\textrm{SU}(3)}}$ Hubbard model,  not only 
mesons (doublons) but also baryons (trions) are naturally generated during the time evolution. 
In the strongly interacting limit, baryons become heavy and attract each other strongly, and their residual 
interaction with mesons generates  meson  diffusion, as captured by the evolution of the equal time density correlation 
function.  Hadrons remain present in the long time limit, while the system  thermalizes to a negative temperature state.
Our conclusions extend to a large variety of initial conditions, all spatial dimensions,  and for SU($N>2$) Hubbard models.

\end{abstract}
\maketitle

\mysection{Introduction}
Non-equilibrium dynamics already provided us with a plethora of interesting phenomena, ranging from negative
temperature states~\cite{braun.2013, Abraham.2017} through universal scaling across quantum phase 
transitions~\cite{Ritsch.2013,Garttner.2017,Heyl.2018,Song.2022} to thermalization dynamics
\cite{Rigol.2008, Jurgen.2021,Langen.2016}.
In this context, recently there has been a large surge of interest in 
understanding and analyzing confinement in spin chains~\cite{Kormos.2017, Surace.2021,Vovrosh.2022,arxiv.2202.12908,arxiv.2206.10528}.
Not only is it related to many body localization  and slow entanglement dynamics~\cite{Bardarson.2012,Huse.2015, Brenes.2018}, 
but confinement is also responsible for creating bound states 
of several interacting particles\cite{Pollack.2009,Greene.2017,Liu.2020}.
Such objects play a prominent role in quantum chromodynamics
and can shed light on particle formation in the early universe~\cite{Alkofer.2001}.


In addition to confinement, short range attractive interactions also give rise to many-particle 
bound states~\cite{Ketterle.2004,Chin.2004}.
These are also responsible for Cooper pairing~\cite{Randeria.1989} and can be engineered in ultracold atomic setting, which
 provide a fantastic platform to simulate and investigate strongly interacting
forms of matter in a laboratory framework~\cite{Bloch.2008, Giorgini.2008,Esslinger.2010, Guan.2013}.
The hyperfine spin of fermionic atoms, in particular,
can play the role of  quark colors or spins in condensed matter, and thereby enable experimentalists
to emulate  phenomena appearing in quantum chromodynamics~\cite{Alford.1998, Alford.2008,Berges.1999}
in the context of hadronic matter as well as simulating $\SUN$ generalizations of condensed matter 
systems~\cite{Honerkamp.2004,He.2006,Rapp.2007,Gorshkov.2010, Scazza.2014,Cazalilla.2014,Hofrichter.2016}.
Attractive fermions such as $^6\textrm{Li}$, e.g., have been
proposed to display color superfluidity and baryon formation 
at sufficiently low temperatures~\cite{Honerkamp.2004,Rapp.2007}.


Fermion gases with attractive interaction are, however, 
quite rare, and  rather unstable against three-body losses, especially in the regime of large scattering 
length~\cite{Huckans.2009}. Moreover, cooling down a fermion gas is a notoriously hard problem~\cite{Fukuhara.2007,Bloch.2008}. 
It is  mostly for these reasons that the state proposed in Ref.~\cite{Rapp.2007} has 
not been observed so far.

\begin{figure*}[t]
\includegraphics[width=0.9\textwidth]{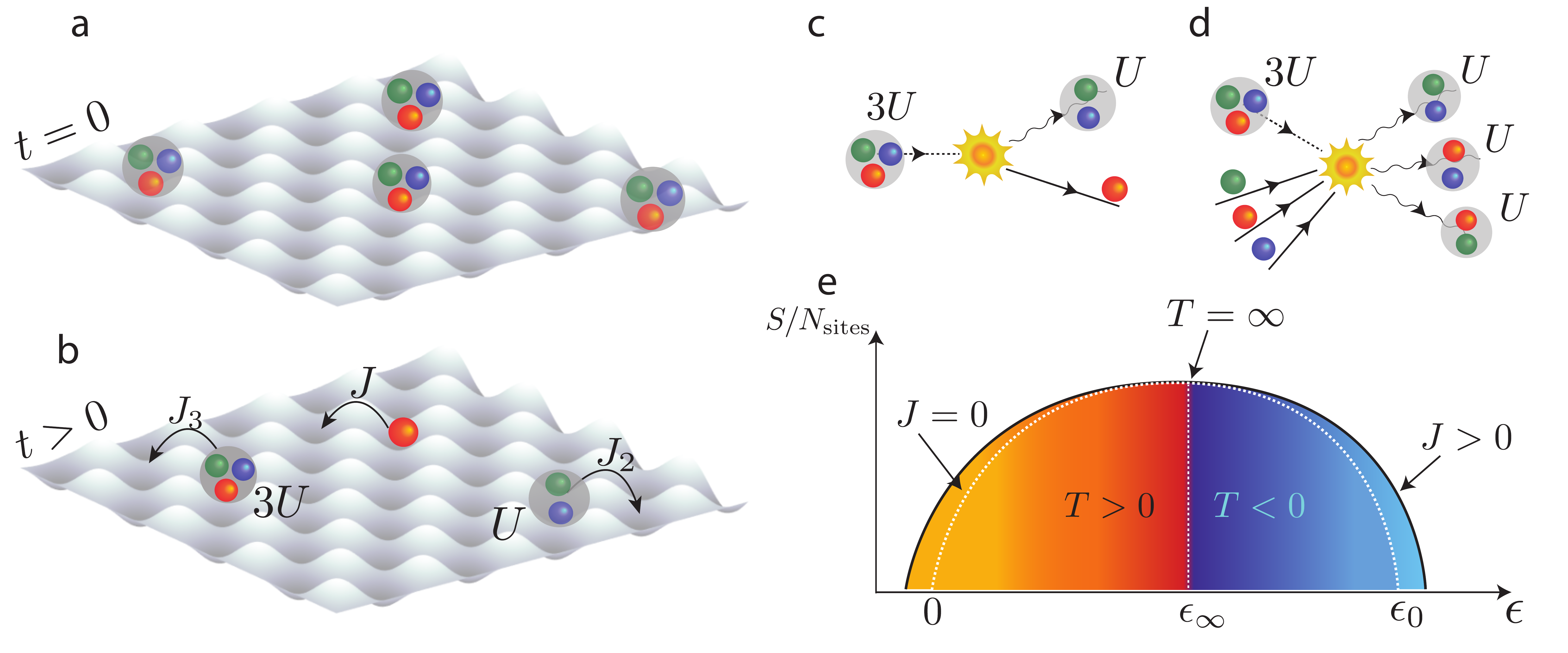}
\caption{\textbf{a} The initial state $|\Xi\rangle$ for $N=3$. Sites are either 
empty or occupied by $N$ fermions of different colors.
\textbf{b}  Schematic motion  for times $t>0$:  $n$-particle composites move
between neighboring sites with hopping  $J_n$, and have an
on-site interaction energy  $n(n-1)/2 U$. \textbf{c} 
Decay of $N$-ions ($N=3$ in the figure) to smaller composites is forbidden by 
energy conservation. \textbf{d}  Lowest order decay channel for $N=3$.  A 
 a trion (baryon) and three fermions (quarks) transform into three doublons (mesons). 
\textbf{e} Sketch of the entropy density as a function of energy density. 
The white dashed line indicates the $J=0$ limit. The entropy density has a 
maximum at energy density $\epsilon_\infty$, where  the temperature diverges
and changes sign. For states with $\epsilon < \epsilon_\infty$ the temperature is positive, 
while for states with $\epsilon > \epsilon_\infty$ it becomes negative. }
\label{fig:sketch}
\end{figure*} 

Due to these complications, these states are not only extremely difficult to create in a controlled fashion, but also their detection is elusive.
As we discuss here, the above problems can be  circumvented and hadron formation naturally appears for fermions 
with \emph{strongly repulsive} interactions, too,  for a set of very simple initial states.  
We also demonstrate that by using quench spectroscopy, we can detect these newly formed particles unambiguously.
This observation opens the door towards simulating and capturing hadron formation with appropriately 
engineered initial state in a much more stable cold atomic environment.  
 Indeed, cold  atomic systems such as $^{137}\text{Yb}$ or $^{87}\text{Sr}$ are today 
 almost routinely used to 
 realize  stable repulsive $\SUN$  gases~\cite{Taie.2010,Taie.2012,Zhang.2014,Sonderhous.2020}.  In an optical lattice, 
 they realize the  repulsive $\SUN$ Hubbard model~\cite{Ulbricht.2010,Dutta.2015},  
 \begin{equation}
 {H} = -J \sum_{\langle \br,\br' \rangle} \sum_{\alpha=1}^N \left( \psi^\dag_{\br \alpha} \psi_{\br'  \alpha} + h.c. \right) + \frac{U}{2} \sum_{\br} n_\br \left( n_\br-1 \right)\;, \label{eq:Hamiltonian}
\end{equation}
describing fermions of $N$ different colors, $\psi_{\br \alpha}^\dagger$, moving around a lattice, and interacting locally 
via a color-independent interaction,  $U$,  
with $n_\br = \sum_\alpha  \psi_{\br \alpha}^\dagger \psi_{\br \alpha}$ the number of fermions at site $\br$. 

The initial state which we propose to engineer is a product of $N$-fermion states~\cite{Winkler.2006,Wang.2010}, 
\begin{equation}
 \left| \Xi \right\rangle = \prod_{\br\; \in \; \Xi} \left( \prod_{\alpha=1}^N \psi^\dag_{\br \alpha} \right) \; \left| 0 \right\rangle \; ,
 \label{eq:Phi_0}
\end{equation}
where  $\Xi$ denotes a subset of sites.   As we show, this state, depicted for $N=3$ in Fig.~\ref{fig:sketch}, 
necessarily evolves to a negative temperature  state for \emph{any} $U>0$, but its dynamical properties change 
dramatically upon increasing the ratio, $U/J$, 
and for $U/J\gtrsim 2$ a strongly interacting quantum gas of baryons and  mesons emerges. 
Similar initial states for $N=2$ have been investigated in Ref. \cite{Trotzky.2012}.

\begin{figure*}[t]
  \includegraphics[width=0.9\textwidth]{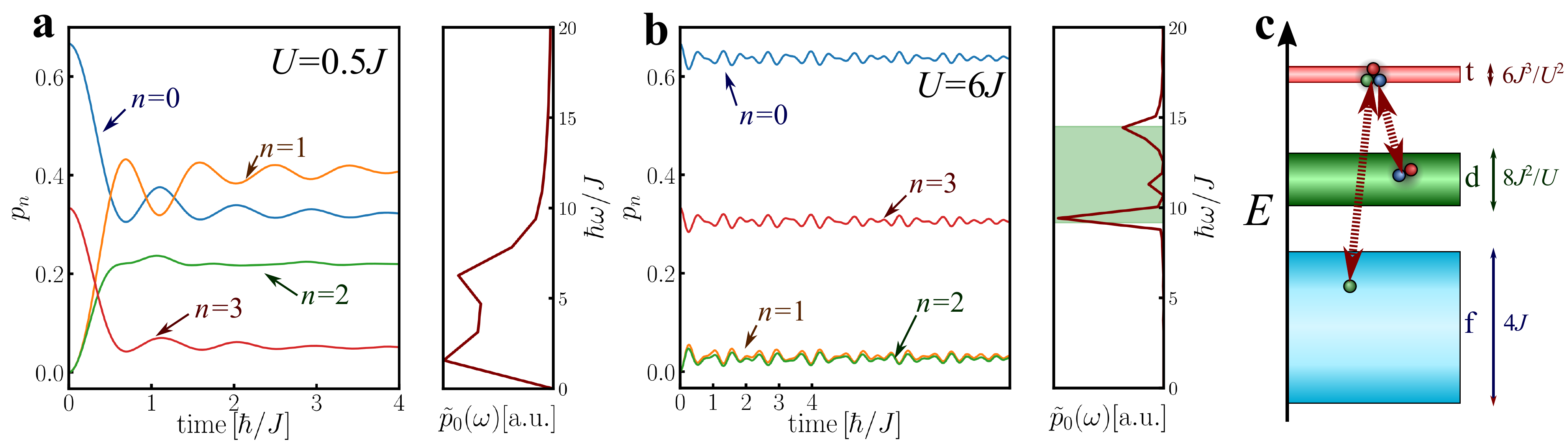}
  \caption{\label{fig:spectroscopy} \textbf{a} 
  Evolution of the probabilities $p_n$ for $U/J=0.5$. 
  The number of empty and triply-occupied sites decreases rapidly, and a large 
  fraction of sites  becomes occupied by one or two fermions. The Fourier spectrum 
  of $p_0(t)$ (right panel) consists of a  spectrum of a broad frequency 
  range up to $\hbar \omega / J \lesssim 10$. 
   \textbf{b} Evolution of the probabilities
  $p_n$ for large interaction strength, $U=6J$. 
  The number of empty and triply-occupied 
   sites remains almost constant apart from small oscillations. The Fourier spectrum of
    $p_0(t)$,  is mostly restricted to  a  window $9.5\lesssim \hbar \omega \lesssim 14.5$. 
    The shading highlights the  predicted spectral window, $\hbar\omega \approx 2U\pm (2J+8J^2/U)$. 
    Frequency peaks at the edges are due to  van Hove singularities  in the single fermion (quark) band.  
         \textbf{c}  Sketch of quasiparticle energies in the large $U$ limit. 
     Oscillation frequencies in panel b can be interpreted as
a result of coherent  oscillations between a baryon  (trion)   and a  state 
      decomposed  into a meson (doublon) and a quark (fermion).}
  \end{figure*}

\mysection{Results}
\\
\mysubsection{Quench spectroscopy and resonances.}
The formation of composite particles can be most easily  detected by quench spectroscopy~\cite{Kormos.2017}, i.e., by analyzing
the time evolution of  the state \eqref{eq:Phi_0} and its Fourier spectrum. We  focused on
 a one-dimensional chain  of $SU(3)$ fermions, and  performed  non-Abelian Time Evolving Block Decimation 
 (TEBD) simulations on it~\cite{Vidal.2004, Vidal.2007, Werner.2020}. On the left panel in Fig.~\ref{fig:spectroscopy}, we show  the time evolution
 of the probabilities $p_n$ of having $n$ fermions at a site for an initial state, where
 three fermions   have been placed at  every third site. 

For $U/J\lesssim 1$,   the probabilities $p_n(t)$ display damped oscillations, and 
the initial occupations $p_3^{(0)} = 1/3$,  $p_0^{(0)} = 2/3$, and 
 $p_1^{(0)} = p_2^{(0)} = 0$ relax to some asymptotic values, $p_n^\infty$.
The Fourier spectra  of $p_n(t)$  consist  of a broad band, consistent with a band of 
width $4J+ U$ of fermionic weakly interacting excitations. 
A careful analysis reveals an exponential relaxation  towards a stationary state with a 
rate $\sim U^2/J/\hbar$ ~\cite{Werner.2020}.

This picture changes radically for $U/J\gtrsim 2$, where the probabilities 
 $p_n(t) $ are pinned roughly to their initial values, indicating that the probability that three 
 particles stay together remains close to 1, $p_3(t)\approx p_3^{(0)}$. The time averaged  probabilities
 of having $n=2$ or $n=1$ fermions at a site  are suppressed, but they are finite, and almost equal. 
Even more strikingly, small, almost undamped oscillations decorate the 
$p_n(t)$ curves. The Fourier trans-
form of these signals reveals
high energy spectral features around  $\hbar\omega \approx 2 U $ (see Fig.~\ref{fig:spectroscopy}b). 
As we demonstrate, the observed  oscillations  
of $ p_n(t) $ can be understood as a result of   \emph{quantum oscillations} 
between \emph{baryonic}  trion states which transform coherently  into \emph{mesonic} doublon 
states and single fermion states,  corresponding to \emph{quarks}.
\mysubsection{Composite particles.}
In the large $U$ limit, we can treat the hopping $J$ as a 
perturbation~\cite{Takahashi.1977,Valmispild.2020}. Isolated 
$n$-particle states have an energy $E_n\approx U n(n-1)/2$ in this limit.
Since the separation $E_n - E_{n-1} = \Delta E_n = (n-1) U \gg J$, these composite 
particles behave as quite stable entities: although  they collide with each other,  they
can decay and transform into each other only under the condition that energy, charge, and  $SU(N)$ spins are 
all conserved.

In the $N=3$ case, e.g., the collision of a baryon (trion) 
of energy $E_3\approx E_3^{(0)}=3U$ with three quarks (free fermions)   of energy $E_1\sim  J\approx 0$ 
and the subsequent decay into  three mesons (doublons)  of energy  $E_2\approx  E_2^{(0)}= U$
provides the lowest order decay channel for baryons~\cite{Strohmaier.2010} (see Fig.~\ref{fig:sketch}d). 
Since quarks have a very small concentration for large $U$,
this event is very unlikely, and baryons 
behave as very stable composite particles. 
They can, however, \emph{virtually} transform back and forth into a 
meson and a quark, and oscillate between these states at a frequency 
$\hbar \omega \approx  E_3^{(0)}- ( E_2^{(0)}+E_1^{(0)}) = 2U$. This process gives rise to the 
oscillations observed  in Fig.~\ref{fig:spectroscopy}b.

Composite particles  move on the lattice with a suppressed hopping. 
 Simple perturbation theory can 
be used to determine  the effective hopping $J_n$ of  an $n$-particle composite, 
yielding  $J_n \approx - n \, (-J)^n \,/ ( (n-1)!\;U^{n-1} ) $ in all dimensions (see Methods). 
Composite  particles are therefore extremely heavy in the large $U$ limit.  In case of $N=3$, the meson band ($n=2$),
is quite narrow compared to the 'quark' band of $\psi_\alpha$ particles, but the baryon band ($n=3$) is even narrower. 
The  spectral peaks  in  Fig.\ref{fig:spectroscopy}b at $\hbar \omega \approx 2 U \pm 2J$ can thus be simply  understood 
as a result of van 
Hove singularities associated with  the edges of the quark band (see Fig.~\ref{fig:spectroscopy}c).

\mysubsection{Effective theory.} 
The emergent composite particles interact very strongly. 
In the large $U$ limit,  heavy $N$-ions dominate, and the density of other composite particles is suppressed. 
The energy of $N$-ions is increased compared to its $J=0$ value by an amount of  $\delta E_N \approx z N \frac{J^2}{(N-1) U}$ 
due to quantum fluctuations, where a  fermion of  color $\alpha$ jumps to one of $z$ neighboring lattice sites. 
Placing two $N$-ions next to each other suppresses these quantum fluctuations, and 
gives rise to an \emph{attractive} interaction,  $V \approx  - 2 N J^2 / ((N-1)\,U)$.

  \begin{figure*}[t]
\includegraphics[width=\textwidth]{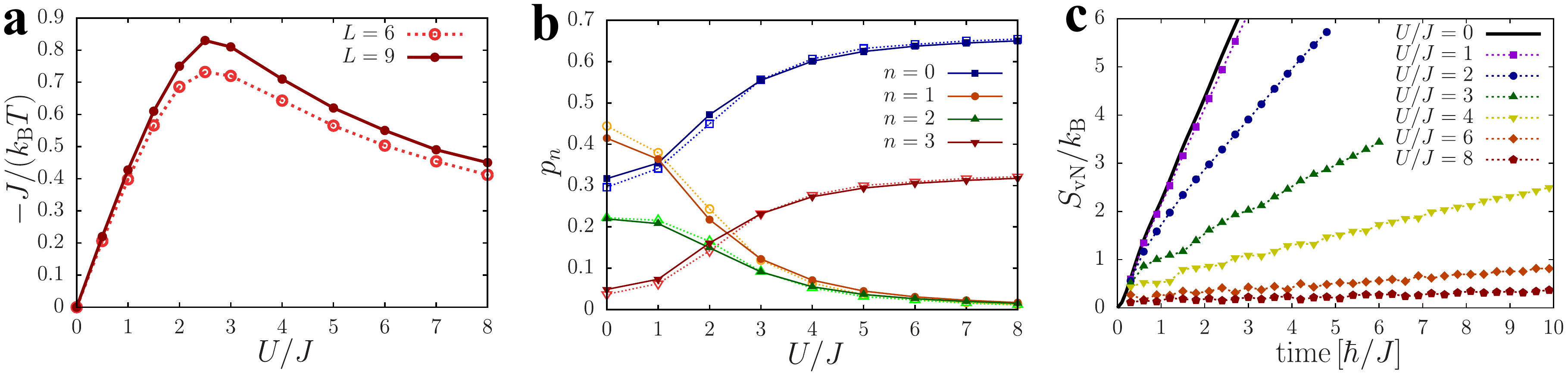}
\caption{\textbf{a} Extracted dimensionless inverse negative temperature $-J/T/k_B$ of the stationary 
state as a function of the interaction strength, $U/J$, as obtained  for small ${{\textrm{SU}(3)}} $ chains 
of lengths $L =  6$ and $L= 9 $ and filling $\nu=1/3$. The fitted value of $-J/T/k_B$ has a clear 
maximum around $U/J \approx 3$. The extracted temperature is negative for any repulsive 
interaction $U$. For small $U$ we obtain $k_B T\approx - 2 J^2/U$, while for large $U$, 
a fall-off $k_B T\propto - U/\ln(U/J)$ is analytically predicted.
\textbf{b} Stationary (long-time) charge distribution $p_n$ 
as a function of the interaction strength. Filled symbols represent data obtained 
from NA-TEBD simulations by calculating the time average of $p_n(t)$ for long times. 
Thermal predictions for a short chain of length $L=9$ are shown by empty symbols. 
\textbf{c} Evolution of the von Neumann entropy of a half chain for different 
interaction strengths. We observe linear growth of entropy for all values of $U$ 
but the growth rate is strongly reduced for large $U$ values. }
\label{fig:entropy}
\end{figure*}

Introducing the operator $\Phi_\br^\dagger$, creating a dressed 
$N$-ion at site  $\br$, we  arrive at the following effective 
 Hamiltonian in  the large $U$ limit
 \begin{eqnarray}
 {H}_{\mathrm{eff}} & = &    -  J_N \sum_{(\br,\br')}  \left(\Phi_\br^\dagger \Phi_{\br'} + h.c. \right) 
 \nonumber
 \\
  &+& E_N \sum_{\br}  \Phi_\br^\dagger \Phi_{\br}-  |V| \sum_{(\br,\br')}   n^\Phi_{\br} n^{\Phi}_{\br'}\;,
  \label{eq:H_trion}
\end{eqnarray}
with $n^{\Phi}_{\br} = \Phi^\dagger_\br \Phi_\br$ the $N$-ion number operator.
Notice that for $N$ odd, the $\Phi$ particles are fermions, while for $N$ even they are hard core bosons.  
Also notice that $|V / J_N| \sim (U/J)^{N-2}$, implying that $N$-ion – $N$-ion interactions become strong for any 
$N>2$ for $U\gg J$. Therefore, in the $U/J\to\infty$ limit,   a gas of spinless, strongly 
interacting $\Phi$ particles is recovered.  For $N=3$, in particular, these particles are dynamically bound fermions, analogous 
to baryons in  QCD.

Although the concentration of other particles is suppressed in the large $U$ limit, their presence is 
still essential. In particular, the concentration $p_{N-1}$ of $(N-1)$-ions  is non-negligible,  $\sim J^2/U^2$. 
 These composite particles, which  transform according to the conjugate representation,  are created 
  by the operators $\Theta^\dagger_{\bar\alpha}$.  Although somewhat lighter than $N$-ions, they are also very
  heavy,  and their interaction  with the $N$-ions is even stronger than the $N$-ion – $N$-ion interaction $V$ itself: 
  neighboring $\Theta_{\;\overline\alpha}$ and 
 $\Phi$ particles can exchange an $\alpha$ fermion in a leading order process, yielding  the effective Hamiltonian
 \begin{equation}
 {H}_{\Phi-\Theta}  \approx     -  J \sum_{(\br,\br')\;,\;\overline \alpha}  
 \bigl(\Phi_\br^\dagger \,\Phi_{\br'} \,\Theta_{\br \overline \alpha}^\dagger\,\Theta_{\br' \overline \alpha} + h.c. \bigr)\;. 
 \end{equation}
It is  this interaction that is ultimately responsible for the motion and transport of  $(N-1)$-ions, which – as we demonstrate later –
dominates mass diffusion on the background of almost immobile  $N$-ions.

\mysubsection{Negative temperature state.} 
We now prove that the initial state  \eqref{eq:Phi_0} must thermalize to a negative temperature state, $T<0$. 
For that we only need to show that the  energy density of the state 
$ | \Xi\rangle$,  
$ \epsilon_0 \equiv    \langle \Xi | {H} | \Xi \rangle / N_\mathrm{sites}  $
is larger than that of the infinite temperature state, 
$ \epsilon_\infty  \equiv    \langle {H}  \rangle_{T=\infty} / N_\mathrm{sites}  $. 
For simplicity, we assume a particle-hole symmetrical band,  but the proof carries over to 
any lattice and any dimension. 
For any product  state of the the form  \eqref{eq:Phi_0},  then one has $ \langle \Xi | {H}_J | \Xi \rangle=0$, and 
one obtains  $\epsilon_0 = \nu \, {U} \, N (N-1) /2$, with $\nu$ the filling factor, i.e. the ratio of sites occupied by $N$-ions. 
Here $H_J$ denotes the hopping part (first term) in Eq. \eqref{eq:Hamiltonian}.

In the infinite temperature limit, the hopping part of the Hamiltonian also averages to zero in case of 
particle-hole symmetrical Hamiltonians. The interaction part can be averaged by observing that each 
$\alpha$ fermion state at a given site is occupied with  probability $\nu$. Thus the  interaction averages to 
 $U \sum_{\alpha<\beta} \langle n_\alpha n_\beta\rangle_{T\to\infty} = {U} \nu^2 \, N (N-1) /2$, yielding 
 $\epsilon_\infty = \nu\,\epsilon_0 < \epsilon_0$. The latter inequality immediately implies 
 that if the state $ | \Xi\rangle$ thermalizes, it must thermalize to a negative temperature 
 state~\cite{Rapp.2010, Rapp.2013,braun.2013} (see Fig.~\ref{fig:sketch}e).
 
 To verify the thermalization of the system~\cite{Deutsch.2018},   we have performed exact diagonalization 
 on small chains of linear sizes $L=9$ and $L=6$, and extracted the effective temperature using the condition 
that the energy density of the thermal state,   $\langle {H}_L  \rangle_{T}/L$,
be equal to  the energy density of the initial state, $\epsilon_0$. We then used the extracted 
negative temperature to evaluate $\langle p_n\rangle_T$, and compared the predicted  
values  with the asymptotic values determined from our simulations. 

Fig.~\ref{fig:entropy}a shows the extracted inverse negative temperatures for both system sizes.  
For small interactions,  a perturbative calculation in $U$ yields   $k_B T_\textrm{eff}\propto - J^2/U$, while for large $U$ we obtain 
 $k_B T_\textrm{eff}\propto - U/\ln(U/J)$, in excellent agreement with the finite size numerical results.
 The probabilities $\langle p_n\rangle_{T_\textrm{eff}}$ agree very well with the dynamically determined values, 
 thereby evidencing the relaxation to a thermal, negative temperature state. 
The extracted negative temperatures  correspond to very ``hot'' states of the system ($|k_B T| \gtrsim J$. 
In such large magnitude negative temperatures, eigenstates with energies much below the upper edge of the 
spectrum are excited. Low energy properties ~\cite{Buchta.2007, Corboz.2012} 
that 
characterize the spectrum near the edge are therefore insufficient to describe the rich dynamics that simulations uncover.

\mysubsection{Entropy growth and  correlations.} 
The emergence of slow, composite particles is clearly visible in the time evolution of von Neumann entropy. 
The initial state at time $t=0$ is a product state, and has vanishing entanglement entropy, wherever we cut the system in 
two. Entropy is generated by particles traveling from one part of the system to another~\cite{Calabrese.2009}. 
As shown in Fig.~\ref{fig:entropy}c,  in our one-dimensional simulations, the von Neumann 
entropy increases linearly with time, and the entropy growth is barely influenced by the interactions as long 
as $U\lesssim J$. For $U\gtrsim J$, however, the entropy growth is rapidly suppressed, indicating that 
particles  carrying the entanglement move very slowly. 
The linear in time entropy growth is expected to be a general features even in non-integrable models, which thermalize 
in the long time limit\cite{Nahum.2020} and do not exhibit many-body localization.

\begin{figure*}[t]
\includegraphics[width=0.9\textwidth]{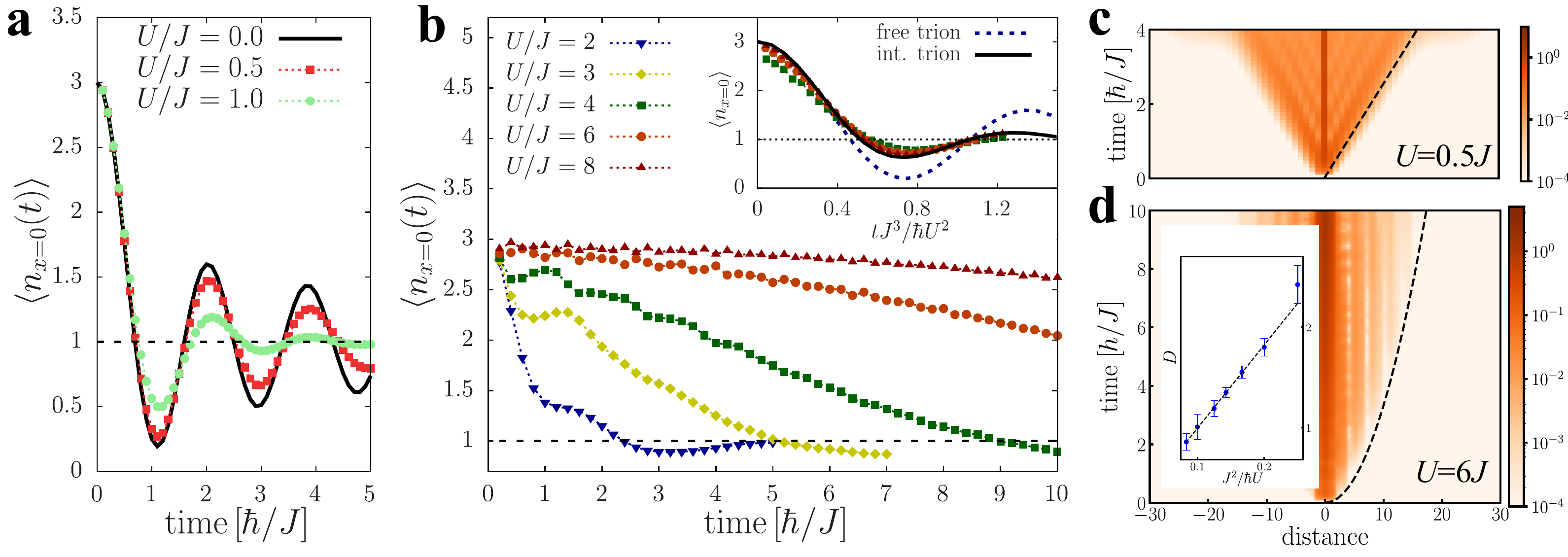}
\caption{\textbf{a} Average fermion number at a triply occupied site as 
a function of time for weak interactions, $U/J \le 1$. In the non-interacting limit, we 
observe algebraically decaying oscillations, which become  weakly but  exponentially damped 
for weak interactions. \textbf{b} 
Strong interactions, $U/J \gtrsim 1$, slow down the dynamics.
Inset: The rescaled curves,  $\langle n_{x=0}(t\,J^3/U^2/\hbar)\rangle $, collapse to a 
single universal curve for $U\to \infty$,  well captured  by  the $ |V| \rightarrow \infty$ limit of the effective
  Hamiltonian \eqref{eq:H_trion} (continuous line). \textbf{c} Amplitude of equal time  density-density 
  correlations $|C_{nn}(x,t)|$ for  $U=0.5J$. Correlations develop
   within a light-cone of a slope $\propto 4J/\hbar$, i.e.,   twice the maximal 
  velocity of free fermions. \textbf{d} Amplitude of equal time  density-density 
  correlations    for strong interaction, $U=6J$. Apart from the central trion peak at $x\approx 0$, 
   correlations spread diffusively according to~\eqref{eq:diffusion}. 
Inset:   The fitted diffusion constant $D$ scales as $D\sim J^2/U$, evidencing meson (doublon) diffusion.
 }
     \label{fig:correlations}
\end{figure*}

The dynamics of composite particles is  more directly captured through time dependent 
charge oscillations in Fig. \ref{fig:correlations}. For small interactions, $U\lesssim J$, the charge at the origin $x=0$, 
$\langle n_{x=0}(t)\rangle$ exhibits weakly damped coherent oscillations with a frequency 
$\sim J$ in Fig. \ref{fig:correlations}a.  This picture changes entirely once we enter the regime $U\gtrsim J$; there
charge oscillations slow down, and universal oscillations with a frequency 
$\hbar\omega\sim J^3/U^2$ appear. Interestingly, these  oscillations are  
\emph{different} from  simple composite fermion oscillations. Rather,  our direct simulations 
with the Hamiltonian \eqref{eq:H_trion} show that for large $U/J$,
$\langle n_0(t)\rangle$ approaches a universal curve described by Eq.~\eqref{eq:H_trion} with 
infinitely strong interaction, $|V| \to \infty$ (see inset of Fig.~ \ref{fig:correlations}b).

Equal time density-density correlations~\cite{Altman.2004}, 
$C_{nn}(x,t)\equiv \langle n_y(t) n_{x+y}(t)\rangle -  \langle n_y(t)\rangle\langle n_{x+y}(t)\rangle$,  
exhibit an even more interesting  picture. In the quark-dominated  $U\lesssim J$ regime, 
a ballistic front propagation and a light-cone structure is observed~\cite{Gopalakrishna.2018} in Fig. \ref{fig:correlations}c. In contrast, for
 $U\gtrsim J$, the  ballistic front is suppressed, and $C_{nn}(x,t)$ consists of two distinct features (see Fig. \ref{fig:correlations}d) .
 A large peak associated with heavy and hardly moving baryons is observed  at $x\approx 0$. 
 In addition,   a diffusive correlation profile appears for $x\not\approx 0$, very well described by the 
expression
\begin{equation}
 |C_{nn}(x,t)| \propto \exp \Bigl( - \frac{x^2}{4 \,t \,D} \Bigr)
 \label{eq:diffusion}
\end{equation}
in one dimension. 
The diffusion constant can be directly extracted from the numerical data, 
and it scales as
\begin{equation}
 D  \propto J^2/U\;.
\end{equation}
This scaling is clearly related to the meson diffusion. 
Mesons have a small and non-negligible  concentration, and move much faster than baryons. 
They collide  with the background trions after a collision time $\tau \sim \hbar U/J^2$, and 
propagate diffusively due to these collisions with a diffusion constant $D  \propto J^2/U/\nu^2$ (see inset of Fig. \ref{fig:correlations}d).

\mysection{Discussion}

{\parindent =0pt In the previous sections} we have shown that hadronic states can be engineered very 
easily in optical lattices: one just has to prepare an initial state, where  $\SUN$ fermions are 
placed in $N$-fermion groups to a subset of lattice sites. 
A repulsive 
interaction $U\gtrsim J$ stabilizes composite particles in this case, 
and bound states and resonances of hadronic nature 
 emerge dynamically as a result of strong interactions. These heavy composite particles are 
simply stabilized by  conservation laws, especially by that of energy and charge conservation, 
and form   a 'hadron' gas of strongly interacting  bosons and  fermions~\cite{Winkler.2006,Wang.2010,Adams.2012}. 

In the particular case of $N=3$, the heaviest hadrons are baryons (trions),   while
composites of two particles behave as somewhat lighter mesons (doublons), in analogy with QCD.  
The original, bare $\psi_\alpha$ fermions  play the role of quarks in this case. 
Similarly rich picture emerges for larger values of $N$, where $n$-ions with $n=1,\dots, N$
exist, and become long-lived particles for $U\gtrsim J$. 

These particles behave somewhat similar to resonances in particle physics: they can transform 
into each other via many-particle collisions, and their densities equilibrate 
with time to form a negative temperature gas of hadrons.  Notice, however, that since the emergent 
composite particles are very heavy and  the decay channels require many-particle collisions, 
thermalization takes place  at time 
scales much longer than the characteristic time $t\sim \hbar/J$ of the bare $\psi_\alpha$ fermions' propagation. 

Interestingly, for bipartite lattices, one can also show that for the   
states $|\Xi\rangle$,  the time evolution of the density of the 
repulsive gas is identical to that of the attractive gas. To prove this, we notice that on a bipartite 
lattice, the gauge transformation ${\cal K}:\;\psi_{\br \alpha}\to (-1)^{||\br||}\psi_{\br\alpha}$,  flipping the sign of the fermion operators 
on one sublattice of the bipartite lattice, changes the sign of $J$, but leaves the  initial state as well as the density operator 
invariant, ${\cal K} :\phantom{n} H_{J,U}\to H_{-J,U} $ and $ |\Xi\rangle \to |\Xi\rangle$.
Combing the time reversal symmetry ${\cal T}$ with ${\cal K}$ then yields, 
${\cal T \,K } \,e^{-i \,t \,H_{J,U}/\hbar}|\Xi\rangle  
=  e^{-i\, t\, H_{J,-U} /\hbar}|\Xi\rangle$. Since both the time reversal and the gauge transformation ${\cal K}$
leave the density invariant, this implies that, starting from the state, $|\Xi\rangle$, the time evolution 
of density correlations is identical for  $U$ and for $-U$. 
Therefore,  the  quantum quench protocol suggested here can also be viewed as effectively realizing
a gas of strongly attractive $\psi$ fermions~\cite{Rapp.2010}. 

Initial states for $N=2$ with two fermions at every second site 
have been realized a long time ago~\cite{Winkler.2006,Wang.2010, Trotzky.2012}.
The preparation of  trion  states could be performed by following similar protocols, but  for $N>2$, 
three particle losses may become an important factor~\cite{Daley.2009,Huckans.2009}. 
To perform a cold atom experiment, one should therefore 
use atoms with a relatively short scattering length $a_s$ compared to the wavelength of the 
optical lattice, $\lambda$. This condition is satisfied by $^{173}\textrm{Yb}$, having a scattering length 
$a_s \approx 10\,\textrm{nm}$~\cite{Hofrichter.2016}, much shorter than the wavelength of the 
confining laser. In fact, a simple calculation yields that the ratio of the interaction $U$ and the three-body loss rate
of a trion, $\gamma$ scales as $U/(\gamma h) \sim (\lambda/a_s)^3$. 
 For the laser used in Ref.~\cite{Hofrichter.2016} with $\lambda = 759 \,\textrm{nm}$, e.g., 
and for a barrier height corresponding to $U/J =1$,   we obtain the estimate  $U/h=J/h\approx 300 \,\textrm{Hz}$, 
while the three-body loss rate remains $\gamma \approx   0.16 \,\textrm{Hz} \ll J/h$. 
The time scale of three-body losses is thus more than $10^3$ times larger than that of the dynamical scale 
of the strongly interacting gas in the regime where baryons and mesons form, 
and should be experimentally accessible~\cite{Hofrichter.2016,Chen.2022}.

Our results demonstrate that not only is it possible to create easily a plethora of interesting particles in the repulsive SU($N>2$) Hubbard model in all dimensions,
but their detection and distinction are also straightforward using quench spectroscopy. This could be useful for many other incarnations of particle production in condensed matter
and cold atomic ensembles.

\mysection{Methods}
\mysubsection{Non-Abelian MPS simulations.}
MPS simulations have been performed for the one dimensional Hubbard model with $N=3$ colors. 
The real time dynamics, generated by the Hamiltonian \eqref{eq:Hamiltonian} 
has been simulated using the infinite chain non-Abelian Time Evolving Block Decimation (NA-TEBD) 
algorithm~\cite{Werner.2020}, while exploiting   the full $\mathrm{SU}(3) \times \mathrm{U}(1)$ symmetry of the model. 
We have kept $M_{\mathrm{mult}} = 2500$   multiplets in the NA-MPS, which
corresponds to a usual bond dimension $M\approx 15000$. The time step in the second order
Suzuki-Trotter approximation was set to $J \Delta t = 0.01$.

We have also performed TEBD simulations using the effective trion Hamiltonian \eqref{eq:H_trion} 
in the $|V| \rightarrow \infty$ limit, by using the residual $\mathrm{U}(1)$ symmetry of the
effective model. Since the infinitely strong interaction forbids trions to hop next to each other, we could map the strongly interacting limit to free fermions by using
a  method of Cheong and Henley \cite{Cheong.2009}.

\mysubsection{Exact Diagonalization at finite T.}
Finite temperature simulations for the $\mathrm{SU}(3)$ Hubbard chain have been 
 carried out for short chains of length $L=6$ and $L=9$ with periodic boundary conditions. 
Corresponding to  $\nu=1/3$ filling,  we have set  the total number of fermions to $n_\mathrm{tot} = L$, 
and also restricted the calculations to the color symmetrical subspace with $n_1 = n_2 = n_3$. 
We have determined the canonical density matrix $\frac{1}{Z} e^{-\beta {H}}$ by using a Taylor expansion. 
The inverse temperature $\beta$ has then been  extracted by enforcing  $\langle\Xi| {H}|\Xi \rangle\equiv \langle{H} \rangle_T $.

\mysubsection{Derivation of effective $N$-ion Hamiltonian.}
The effective Hamiltonians describing the motion of composite particles can
be constructed by performing perturbation theory in $J$. The effective hopping for an $n$-ion, e.g., can be 
estimated in leading order~\cite{Takahashi.1977,MacDonald.1988,Cai.2021} as
$$
- J_n = \langle 0|
\prod_{\alpha = 1}^n \psi_{\br\alpha } \;  T_{\br,\br'} \left(\frac{1}{E_n^{(0)} - H_U}  T _{\br,\br'} \right)^{n-1} \prod_{\beta = 1}^n \psi^\dagger_{\br'\beta } 
 |0\rangle 
$$
where $\br$ and $\br'$ are two neighboring sites, $H_U$ denotes the on site energy, and 
$  T _{\br,\br'} = -J \sum_{\alpha=1}^N \psi^\dagger_{\br\alpha} \psi_{\br'\alpha}$ the hopping operator from   $\br'$ and $\br$. 
In the intermediate state, where $l$ particles are on $\br$ and $(n-l)$ at site $\br'$ the denominator can be evaluated 
as $E_n^{(0)}-E_l^{(0)}-E_{n-l}^{(0)} = U n (n-l)$. The product of the denominators can therefore be evaluated 
as $U \;((n-1)!)^2 $, while the order in which particles are 
transferred between the two sites amounts in a combinatorial  factor, $n!$, yielding the result in the main text.

\mysubsection{Estimation of effective temperature.}

{\parindent = 0pt \emph{Small $U$ limit.} }To estimate the effective temperature for $U\ll J$, we first notice that $\beta$ in this limit is 
a very small negative number. We can therefore perform a perturbative calculation in this limit by keeping only the leading order terms 
in $\beta$ and  $U$, and by taking the expectation  value  $\langle H_J+H_U\rangle_T$ with the unperturbed density operator, 
and equating that with the energy of the initial state. This procedure yields the equation
\be
- \nu(1-\nu) \beta \int \mathrm{d}\epsilon\;   \epsilon\,\rho(\epsilon)  + U \textstyle{\frac {N(N-1)}{2}} \nu^2
\approx \nu \;U \,\frac {N(N-1)}{2}\;,
\label{eq:T_eff}
\ee
where $\rho(\epsilon) $ denotes the single particle density of states of the lattice.  The prefactor $(1-\nu) \nu$ 
 in the  first integral is  associated with the (diverging)  chemical potential, $e^{-\beta\mu}\approx (1-\nu)/\nu$. 
 On a $D$ dimensional cubic lattice Eq.~\eqref{eq:T_eff}  yields
\be
\beta_\textrm{eff} (U\ll J) \approx -  \frac{ (N-1)}{4 D}\, \frac{U}{J^2} \;. 
\ee
In $D=1$ dimension and for  $N=3$ and $\nu=1/3$, we obtain a prefactor $1/2$, in perfect agreement 
with Fig.~\ref{fig:entropy}a.

{\parindent = 0pt \emph{Large $U$ limit.} }In the $U\gg J$ limit, we can consider each site as a grand canonical 
subsystem, weakly coupled to the rest of the system, and having a density matrix $\rho \propto e^{-\beta(U n (n-1)/2 - \mu n)}$. 
Since $N$-ions are abundant, the chemical potential 
of the subsystem must be approximately $\mu\approx U (N-1)/2$. The probability of having a single fermion 
is then given by $p_{n=1} \approx (1-\nu) N e^{\beta U (N-1)/2} $, where the prefactor approximates 
the inverse partition function. 

The value of  $p_{n=1} $ can, however, be also estimated quite simply by noticing that the state, where we have 
$N$ particles at one site, say at the origin, $|N\rangle$, is not a true $N$-ion state at finite $J/U$. Quantum fluctuations 
allow each fermion  forming the $N$-ion to move to neighboring sites with probability amplitudes $\approx J/((N-1) U)$. 
This implies that a state $|N\rangle$ decomposes to a fermion and an $(N-1)\,$-ion with  a probability 
$P_{N\to\; (N-1) \textrm{+} 1} \approx    z N {J^2}/ [(N-1)^2 U^2]$, yielding 
$p_{1} \approx p_{N-1} \approx  z \,\nu \,N {J^2} /[{(N-1)^2 U^2}]$. 
Comparison with the thermal average  then yields 
\be
\beta_\textrm{eff} (U\gg J) \approx -  \frac{4}{(N-1) \,U}\, \ln \Big( \sqrt{\frac{1-\nu}{\nu\, z}}\,\frac {(N-1) U} J \Big)\;,
\ee
with $z=2D$ on a $D$-dimensional cubic lattice. 

\mysubsection{Data availability.}
The data that support the findings of this study are available from the corresponding author upon reasonable request.

\bibliographystyle{apsrev4-2}
\bibliography{bibl}

\begin{thebibliography}{70}%
\makeatletter
\providecommand \@ifxundefined [1]{%
 \@ifx{#1\undefined}
}%
\providecommand \@ifnum [1]{%
 \ifnum #1\expandafter \@firstoftwo
 \else \expandafter \@secondoftwo
 \fi
}%
\providecommand \@ifx [1]{%
 \ifx #1\expandafter \@firstoftwo
 \else \expandafter \@secondoftwo
 \fi
}%
\providecommand \natexlab [1]{#1}%
\providecommand \enquote  [1]{``#1''}%
\providecommand \bibnamefont  [1]{#1}%
\providecommand \bibfnamefont [1]{#1}%
\providecommand \citenamefont [1]{#1}%
\providecommand \href@noop [0]{\@secondoftwo}%
\providecommand \href [0]{\begingroup \@sanitize@url \@href}%
\providecommand \@href[1]{\@@startlink{#1}\@@href}%
\providecommand \@@href[1]{\endgroup#1\@@endlink}%
\providecommand \@sanitize@url [0]{\catcode `\\12\catcode `\$12\catcode
  `\&12\catcode `\#12\catcode `\^12\catcode `\_12\catcode `\%12\relax}%
\providecommand \@@startlink[1]{}%
\providecommand \@@endlink[0]{}%
\providecommand \url  [0]{\begingroup\@sanitize@url \@url }%
\providecommand \@url [1]{\endgroup\@href {#1}{\urlprefix }}%
\providecommand \urlprefix  [0]{URL }%
\providecommand \Eprint [0]{\href }%
\providecommand \doibase [0]{https://doi.org/}%
\providecommand \selectlanguage [0]{\@gobble}%
\providecommand \bibinfo  [0]{\@secondoftwo}%
\providecommand \bibfield  [0]{\@secondoftwo}%
\providecommand \translation [1]{[#1]}%
\providecommand \BibitemOpen [0]{}%
\providecommand \bibitemStop [0]{}%
\providecommand \bibitemNoStop [0]{.\EOS\space}%
\providecommand \EOS [0]{\spacefactor3000\relax}%
\providecommand \BibitemShut  [1]{\csname bibitem#1\endcsname}%
\let\auto@bib@innerbib\@empty
\bibitem [{\citenamefont {Braun}\ \emph {et~al.}(2013)\citenamefont {Braun},
  \citenamefont {Ronzheimer}, \citenamefont {Schreiber}, \citenamefont
  {Hodgman}, \citenamefont {Rom}, \citenamefont {Bloch},\ and\ \citenamefont
  {Schneider}}]{braun.2013}%
  \BibitemOpen
  \bibfield  {author} {\bibinfo {author} {\bibfnamefont {S.}~\bibnamefont
  {Braun}}, \bibinfo {author} {\bibfnamefont {J.~P.}\ \bibnamefont
  {Ronzheimer}}, \bibinfo {author} {\bibfnamefont {M.}~\bibnamefont
  {Schreiber}}, \bibinfo {author} {\bibfnamefont {S.~S.}\ \bibnamefont
  {Hodgman}}, \bibinfo {author} {\bibfnamefont {T.}~\bibnamefont {Rom}},
  \bibinfo {author} {\bibfnamefont {I.}~\bibnamefont {Bloch}},\ and\ \bibinfo
  {author} {\bibfnamefont {U.}~\bibnamefont {Schneider}},\ }\href
  {https://doi.org/10.1126/science.1227831} {\bibfield  {journal} {\bibinfo
  {journal} {Science}\ }\textbf {\bibinfo {volume} {339}},\ \bibinfo {pages}
  {52} (\bibinfo {year} {2013})}\BibitemShut {NoStop}%
\bibitem [{\citenamefont {Abraham}\ and\ \citenamefont
  {Penrose}(2017)}]{Abraham.2017}%
  \BibitemOpen
  \bibfield  {author} {\bibinfo {author} {\bibfnamefont {E.}~\bibnamefont
  {Abraham}}\ and\ \bibinfo {author} {\bibfnamefont {O.}~\bibnamefont
  {Penrose}},\ }\href {https://doi.org/10.1103/PhysRevE.95.012125} {\bibfield
  {journal} {\bibinfo  {journal} {Phys. Rev. E}\ }\textbf {\bibinfo {volume}
  {95}},\ \bibinfo {pages} {012125} (\bibinfo {year} {2017})}\BibitemShut
  {NoStop}%
\bibitem [{\citenamefont {Ritsch}\ \emph {et~al.}(2013)\citenamefont {Ritsch},
  \citenamefont {Domokos}, \citenamefont {Brennecke},\ and\ \citenamefont
  {Esslinger}}]{Ritsch.2013}%
  \BibitemOpen
  \bibfield  {author} {\bibinfo {author} {\bibfnamefont {H.}~\bibnamefont
  {Ritsch}}, \bibinfo {author} {\bibfnamefont {P.}~\bibnamefont {Domokos}},
  \bibinfo {author} {\bibfnamefont {F.}~\bibnamefont {Brennecke}},\ and\
  \bibinfo {author} {\bibfnamefont {T.}~\bibnamefont {Esslinger}},\ }\href
  {https://doi.org/10.1103/RevModPhys.85.553} {\bibfield  {journal} {\bibinfo
  {journal} {Rev. Mod. Phys.}\ }\textbf {\bibinfo {volume} {85}},\ \bibinfo
  {pages} {553} (\bibinfo {year} {2013})}\BibitemShut {NoStop}%
\bibitem [{\citenamefont {G{\"a}rttner}\ \emph {et~al.}(2017)\citenamefont
  {G{\"a}rttner}, \citenamefont {Bohnet}, \citenamefont {Safavi-Naini},
  \citenamefont {Wall}, \citenamefont {Bollinger},\ and\ \citenamefont
  {Rey}}]{Garttner.2017}%
  \BibitemOpen
  \bibfield  {author} {\bibinfo {author} {\bibfnamefont {M.}~\bibnamefont
  {G{\"a}rttner}}, \bibinfo {author} {\bibfnamefont {J.~G.}\ \bibnamefont
  {Bohnet}}, \bibinfo {author} {\bibfnamefont {A.}~\bibnamefont
  {Safavi-Naini}}, \bibinfo {author} {\bibfnamefont {M.~L.}\ \bibnamefont
  {Wall}}, \bibinfo {author} {\bibfnamefont {J.~J.}\ \bibnamefont
  {Bollinger}},\ and\ \bibinfo {author} {\bibfnamefont {A.~M.}\ \bibnamefont
  {Rey}},\ }\href {https://doi.org/10.1038/nphys4119} {\bibfield  {journal}
  {\bibinfo  {journal} {Nature Physics}\ }\textbf {\bibinfo {volume} {13}},\
  \bibinfo {pages} {781} (\bibinfo {year} {2017})}\BibitemShut {NoStop}%
\bibitem [{\citenamefont {Heyl}(2018)}]{Heyl.2018}%
  \BibitemOpen
  \bibfield  {author} {\bibinfo {author} {\bibfnamefont {M.}~\bibnamefont
  {Heyl}},\ }\href {https://doi.org/10.1088/1361-6633/aaaf9a} {\bibfield
  {journal} {\bibinfo  {journal} {Reports on Progress in Physics}\ }\textbf
  {\bibinfo {volume} {81}},\ \bibinfo {pages} {054001} (\bibinfo {year}
  {2018})}\BibitemShut {NoStop}%
\bibitem [{\citenamefont {Song}\ \emph {et~al.}(2022)\citenamefont {Song},
  \citenamefont {Dutta}, \citenamefont {Bhave}, \citenamefont {Yu},
  \citenamefont {Carter}, \citenamefont {Cooper},\ and\ \citenamefont
  {Schneider}}]{Song.2022}%
  \BibitemOpen
  \bibfield  {author} {\bibinfo {author} {\bibfnamefont {B.}~\bibnamefont
  {Song}}, \bibinfo {author} {\bibfnamefont {S.}~\bibnamefont {Dutta}},
  \bibinfo {author} {\bibfnamefont {S.}~\bibnamefont {Bhave}}, \bibinfo
  {author} {\bibfnamefont {J.-C.}\ \bibnamefont {Yu}}, \bibinfo {author}
  {\bibfnamefont {E.}~\bibnamefont {Carter}}, \bibinfo {author} {\bibfnamefont
  {N.}~\bibnamefont {Cooper}},\ and\ \bibinfo {author} {\bibfnamefont
  {U.}~\bibnamefont {Schneider}},\ }\href
  {https://doi.org/10.1038/s41567-021-01476-w} {\bibfield  {journal} {\bibinfo
  {journal} {Nature Physics}\ }\textbf {\bibinfo {volume} {18}},\ \bibinfo
  {pages} {259} (\bibinfo {year} {2022})}\BibitemShut {NoStop}%
\bibitem [{\citenamefont {Rigol}\ \emph {et~al.}(2008)\citenamefont {Rigol},
  \citenamefont {Dunjko},\ and\ \citenamefont {Olshanii}}]{Rigol.2008}%
  \BibitemOpen
  \bibfield  {author} {\bibinfo {author} {\bibfnamefont {M.}~\bibnamefont
  {Rigol}}, \bibinfo {author} {\bibfnamefont {V.}~\bibnamefont {Dunjko}},\ and\
  \bibinfo {author} {\bibfnamefont {M.}~\bibnamefont {Olshanii}},\ }\href
  {https://doi.org/10.1038/nature06838} {\bibfield  {journal} {\bibinfo
  {journal} {Nature}\ }\textbf {\bibinfo {volume} {452}},\ \bibinfo {pages}
  {854} (\bibinfo {year} {2008})}\BibitemShut {NoStop}%
\bibitem [{\citenamefont {Berges}\ \emph {et~al.}(2021)\citenamefont {Berges},
  \citenamefont {Heller}, \citenamefont {Mazeliauskas},\ and\ \citenamefont
  {Venugopalan}}]{Jurgen.2021}%
  \BibitemOpen
  \bibfield  {author} {\bibinfo {author} {\bibfnamefont {J.}~\bibnamefont
  {Berges}}, \bibinfo {author} {\bibfnamefont {M.~P.}\ \bibnamefont {Heller}},
  \bibinfo {author} {\bibfnamefont {A.}~\bibnamefont {Mazeliauskas}},\ and\
  \bibinfo {author} {\bibfnamefont {R.}~\bibnamefont {Venugopalan}},\ }\href
  {https://doi.org/10.1103/RevModPhys.93.035003} {\bibfield  {journal}
  {\bibinfo  {journal} {Rev. Mod. Phys.}\ }\textbf {\bibinfo {volume} {93}},\
  \bibinfo {pages} {035003} (\bibinfo {year} {2021})}\BibitemShut {NoStop}%
\bibitem [{\citenamefont {Langen}\ \emph {et~al.}(2016)\citenamefont {Langen},
  \citenamefont {Gasenzer},\ and\ \citenamefont {Schmiedmayer}}]{Langen.2016}%
  \BibitemOpen
  \bibfield  {author} {\bibinfo {author} {\bibfnamefont {T.}~\bibnamefont
  {Langen}}, \bibinfo {author} {\bibfnamefont {T.}~\bibnamefont {Gasenzer}},\
  and\ \bibinfo {author} {\bibfnamefont {J.}~\bibnamefont {Schmiedmayer}},\
  }\href {https://doi.org/10.1088/1742-5468/2016/06/064009} {\bibfield
  {journal} {\bibinfo  {journal} {Journal of Statistical Mechanics: Theory and
  Experiment}\ }\textbf {\bibinfo {volume} {2016}},\ \bibinfo {pages} {064009}
  (\bibinfo {year} {2016})}\BibitemShut {NoStop}%
\bibitem [{\citenamefont {Kormos}\ \emph {et~al.}(2017)\citenamefont {Kormos},
  \citenamefont {Collura}, \citenamefont {Tak{\'a}cs},\ and\ \citenamefont
  {Calabrese}}]{Kormos.2017}%
  \BibitemOpen
  \bibfield  {author} {\bibinfo {author} {\bibfnamefont {M.}~\bibnamefont
  {Kormos}}, \bibinfo {author} {\bibfnamefont {M.}~\bibnamefont {Collura}},
  \bibinfo {author} {\bibfnamefont {G.}~\bibnamefont {Tak{\'a}cs}},\ and\
  \bibinfo {author} {\bibfnamefont {P.}~\bibnamefont {Calabrese}},\ }\href
  {https://doi.org/10.1038/nphys3934} {\bibfield  {journal} {\bibinfo
  {journal} {Nature Physics}\ }\textbf {\bibinfo {volume} {13}},\ \bibinfo
  {pages} {246} (\bibinfo {year} {2017})}\BibitemShut {NoStop}%
\bibitem [{\citenamefont {Surace}\ and\ \citenamefont
  {Lerose}(2021)}]{Surace.2021}%
  \BibitemOpen
  \bibfield  {author} {\bibinfo {author} {\bibfnamefont {F.~M.}\ \bibnamefont
  {Surace}}\ and\ \bibinfo {author} {\bibfnamefont {A.}~\bibnamefont
  {Lerose}},\ }\href {https://doi.org/10.1088/1367-2630/abfc40} {\bibfield
  {journal} {\bibinfo  {journal} {New Journal of Physics}\ }\textbf {\bibinfo
  {volume} {23}},\ \bibinfo {pages} {062001} (\bibinfo {year}
  {2021})}\BibitemShut {NoStop}%
\bibitem [{\citenamefont {Vovrosh}\ \emph {et~al.}(2022)\citenamefont
  {Vovrosh}, \citenamefont {Mukherjee}, \citenamefont {Bastianello},\ and\
  \citenamefont {Knolle}}]{Vovrosh.2022}%
  \BibitemOpen
  \bibfield  {author} {\bibinfo {author} {\bibfnamefont {J.}~\bibnamefont
  {Vovrosh}}, \bibinfo {author} {\bibfnamefont {R.}~\bibnamefont {Mukherjee}},
  \bibinfo {author} {\bibfnamefont {A.}~\bibnamefont {Bastianello}},\ and\
  \bibinfo {author} {\bibfnamefont {J.}~\bibnamefont {Knolle}},\ }\href@noop {}
  {\bibinfo {title} {Dynamical hadron formation in long-range interacting
  quantum spin chains}} (\bibinfo {year} {2022}),\ \Eprint
  {https://arxiv.org/abs/2204.05641} {arXiv:2204.05641 [cond-mat.quant-gas]}
  \BibitemShut {NoStop}%
\bibitem [{\citenamefont {Birnkammer}\ \emph {et~al.}(2022)\citenamefont
  {Birnkammer}, \citenamefont {Bastianello},\ and\ \citenamefont
  {Knap}}]{arxiv.2202.12908}%
  \BibitemOpen
  \bibfield  {author} {\bibinfo {author} {\bibfnamefont {S.}~\bibnamefont
  {Birnkammer}}, \bibinfo {author} {\bibfnamefont {A.}~\bibnamefont
  {Bastianello}},\ and\ \bibinfo {author} {\bibfnamefont {M.}~\bibnamefont
  {Knap}},\ }\href {https://doi.org/10.48550/ARXIV.2202.12908} {\bibinfo
  {title} {Prethermalization in confined spin chains}} (\bibinfo {year}
  {2022}),\ \Eprint {https://arxiv.org/abs/2202.12908} {arXiv:2202.12908}
  \BibitemShut {NoStop}%
\bibitem [{\citenamefont {Banuls}\ \emph {et~al.}(2022)\citenamefont {Banuls},
  \citenamefont {Heller}, \citenamefont {Jansen}, \citenamefont {Knaute},\ and\
  \citenamefont {Svensson}}]{arxiv.2206.10528}%
  \BibitemOpen
  \bibfield  {author} {\bibinfo {author} {\bibfnamefont {M.~C.}\ \bibnamefont
  {Banuls}}, \bibinfo {author} {\bibfnamefont {M.~P.}\ \bibnamefont {Heller}},
  \bibinfo {author} {\bibfnamefont {K.}~\bibnamefont {Jansen}}, \bibinfo
  {author} {\bibfnamefont {J.}~\bibnamefont {Knaute}},\ and\ \bibinfo {author}
  {\bibfnamefont {V.}~\bibnamefont {Svensson}},\ }\href
  {https://doi.org/10.48550/ARXIV.2206.10528} {\bibinfo {title} {A quantum
  information perspective on meson melting}} (\bibinfo {year} {2022}),\ \Eprint
  {https://arxiv.org/abs/2206.10528} {arXiv:2206.10528} \BibitemShut {NoStop}%
\bibitem [{\citenamefont {Bardarson}\ \emph {et~al.}(2012)\citenamefont
  {Bardarson}, \citenamefont {Pollmann},\ and\ \citenamefont
  {Moore}}]{Bardarson.2012}%
  \BibitemOpen
  \bibfield  {author} {\bibinfo {author} {\bibfnamefont {J.~H.}\ \bibnamefont
  {Bardarson}}, \bibinfo {author} {\bibfnamefont {F.}~\bibnamefont
  {Pollmann}},\ and\ \bibinfo {author} {\bibfnamefont {J.~E.}\ \bibnamefont
  {Moore}},\ }\href {https://doi.org/10.1103/PhysRevLett.109.017202} {\bibfield
   {journal} {\bibinfo  {journal} {Phys. Rev. Lett.}\ }\textbf {\bibinfo
  {volume} {109}},\ \bibinfo {pages} {017202} (\bibinfo {year}
  {2012})}\BibitemShut {NoStop}%
\bibitem [{\citenamefont {Nandkishore}\ and\ \citenamefont
  {Huse}(2015)}]{Huse.2015}%
  \BibitemOpen
  \bibfield  {author} {\bibinfo {author} {\bibfnamefont {R.}~\bibnamefont
  {Nandkishore}}\ and\ \bibinfo {author} {\bibfnamefont {D.~A.}\ \bibnamefont
  {Huse}},\ }\href {https://doi.org/10.1146/annurev-conmatphys-031214-014726}
  {\bibfield  {journal} {\bibinfo  {journal} {Annual Review of Condensed Matter
  Physics}\ }\textbf {\bibinfo {volume} {6}},\ \bibinfo {pages} {15} (\bibinfo
  {year} {2015})}\BibitemShut {NoStop}%
\bibitem [{\citenamefont {Brenes}\ \emph {et~al.}(2018)\citenamefont {Brenes},
  \citenamefont {Dalmonte}, \citenamefont {Heyl},\ and\ \citenamefont
  {Scardicchio}}]{Brenes.2018}%
  \BibitemOpen
  \bibfield  {author} {\bibinfo {author} {\bibfnamefont {M.}~\bibnamefont
  {Brenes}}, \bibinfo {author} {\bibfnamefont {M.}~\bibnamefont {Dalmonte}},
  \bibinfo {author} {\bibfnamefont {M.}~\bibnamefont {Heyl}},\ and\ \bibinfo
  {author} {\bibfnamefont {A.}~\bibnamefont {Scardicchio}},\ }\href
  {https://doi.org/10.1103/PhysRevLett.120.030601} {\bibfield  {journal}
  {\bibinfo  {journal} {Phys. Rev. Lett.}\ }\textbf {\bibinfo {volume} {120}},\
  \bibinfo {pages} {030601} (\bibinfo {year} {2018})}\BibitemShut {NoStop}%
\bibitem [{\citenamefont {Pollack}\ \emph {et~al.}(2009)\citenamefont
  {Pollack}, \citenamefont {Dries},\ and\ \citenamefont
  {Hulet}}]{Pollack.2009}%
  \BibitemOpen
  \bibfield  {author} {\bibinfo {author} {\bibfnamefont {S.~E.}\ \bibnamefont
  {Pollack}}, \bibinfo {author} {\bibfnamefont {D.}~\bibnamefont {Dries}},\
  and\ \bibinfo {author} {\bibfnamefont {R.~G.}\ \bibnamefont {Hulet}},\ }\href
  {https://doi.org/10.1126/science.1182840} {\bibfield  {journal} {\bibinfo
  {journal} {Science}\ }\textbf {\bibinfo {volume} {326}},\ \bibinfo {pages}
  {1683} (\bibinfo {year} {2009})}\BibitemShut {NoStop}%
\bibitem [{\citenamefont {Greene}\ \emph {et~al.}(2017)\citenamefont {Greene},
  \citenamefont {Giannakeas},\ and\ \citenamefont
  {P\'erez-R\'{\i}os}}]{Greene.2017}%
  \BibitemOpen
  \bibfield  {author} {\bibinfo {author} {\bibfnamefont {C.~H.}\ \bibnamefont
  {Greene}}, \bibinfo {author} {\bibfnamefont {P.}~\bibnamefont {Giannakeas}},\
  and\ \bibinfo {author} {\bibfnamefont {J.}~\bibnamefont
  {P\'erez-R\'{\i}os}},\ }\href {https://doi.org/10.1103/RevModPhys.89.035006}
  {\bibfield  {journal} {\bibinfo  {journal} {Rev. Mod. Phys.}\ }\textbf
  {\bibinfo {volume} {89}},\ \bibinfo {pages} {035006} (\bibinfo {year}
  {2017})}\BibitemShut {NoStop}%
\bibitem [{\citenamefont {Liu}\ \emph {et~al.}(2020)\citenamefont {Liu},
  \citenamefont {Whitsitt}, \citenamefont {Bienias}, \citenamefont {Lundgren},\
  and\ \citenamefont {Gorshkov}}]{Liu.2020}%
  \BibitemOpen
  \bibfield  {author} {\bibinfo {author} {\bibfnamefont {F.}~\bibnamefont
  {Liu}}, \bibinfo {author} {\bibfnamefont {S.}~\bibnamefont {Whitsitt}},
  \bibinfo {author} {\bibfnamefont {P.}~\bibnamefont {Bienias}}, \bibinfo
  {author} {\bibfnamefont {R.}~\bibnamefont {Lundgren}},\ and\ \bibinfo
  {author} {\bibfnamefont {A.~V.}\ \bibnamefont {Gorshkov}},\ }\href
  {https://doi.org/10.48550/arXiv.2007.07258} {\bibinfo {title} {Realizing and
  probing baryonic excitations in rydberg atom arrays}} (\bibinfo {year}
  {2020}),\ \Eprint {https://arxiv.org/abs/2007.07258} {arXiv:2007.07258}
  \BibitemShut {NoStop}%
\bibitem [{\citenamefont {Alkofer}\ and\ \citenamefont {{von
  Smekal}}(2001)}]{Alkofer.2001}%
  \BibitemOpen
  \bibfield  {author} {\bibinfo {author} {\bibfnamefont {R.}~\bibnamefont
  {Alkofer}}\ and\ \bibinfo {author} {\bibfnamefont {L.}~\bibnamefont {{von
  Smekal}}},\ }\href
  {https://doi.org/https://doi.org/10.1016/S0370-1573(01)00010-2} {\bibfield
  {journal} {\bibinfo  {journal} {Physics Reports}\ }\textbf {\bibinfo {volume}
  {353}},\ \bibinfo {pages} {281} (\bibinfo {year} {2001})}\BibitemShut
  {NoStop}%
\bibitem [{\citenamefont {Zwierlein}\ \emph {et~al.}(2004)\citenamefont
  {Zwierlein}, \citenamefont {Stan}, \citenamefont {Schunck}, \citenamefont
  {Raupach}, \citenamefont {Kerman},\ and\ \citenamefont
  {Ketterle}}]{Ketterle.2004}%
  \BibitemOpen
  \bibfield  {author} {\bibinfo {author} {\bibfnamefont {M.~W.}\ \bibnamefont
  {Zwierlein}}, \bibinfo {author} {\bibfnamefont {C.~A.}\ \bibnamefont {Stan}},
  \bibinfo {author} {\bibfnamefont {C.~H.}\ \bibnamefont {Schunck}}, \bibinfo
  {author} {\bibfnamefont {S.~M.~F.}\ \bibnamefont {Raupach}}, \bibinfo
  {author} {\bibfnamefont {A.~J.}\ \bibnamefont {Kerman}},\ and\ \bibinfo
  {author} {\bibfnamefont {W.}~\bibnamefont {Ketterle}},\ }\href
  {https://doi.org/10.1103/PhysRevLett.92.120403} {\bibfield  {journal}
  {\bibinfo  {journal} {Phys. Rev. Lett.}\ }\textbf {\bibinfo {volume} {92}},\
  \bibinfo {pages} {120403} (\bibinfo {year} {2004})}\BibitemShut {NoStop}%
\bibitem [{\citenamefont {Chin}\ \emph {et~al.}(2004)\citenamefont {Chin},
  \citenamefont {Bartenstein}, \citenamefont {Altmeyer}, \citenamefont {Riedl},
  \citenamefont {Jochim}, \citenamefont {Denschlag},\ and\ \citenamefont
  {Grimm}}]{Chin.2004}%
  \BibitemOpen
  \bibfield  {author} {\bibinfo {author} {\bibfnamefont {C.}~\bibnamefont
  {Chin}}, \bibinfo {author} {\bibfnamefont {M.}~\bibnamefont {Bartenstein}},
  \bibinfo {author} {\bibfnamefont {A.}~\bibnamefont {Altmeyer}}, \bibinfo
  {author} {\bibfnamefont {S.}~\bibnamefont {Riedl}}, \bibinfo {author}
  {\bibfnamefont {S.}~\bibnamefont {Jochim}}, \bibinfo {author} {\bibfnamefont
  {J.~H.}\ \bibnamefont {Denschlag}},\ and\ \bibinfo {author} {\bibfnamefont
  {R.}~\bibnamefont {Grimm}},\ }\href {https://doi.org/10.1126/science.1100818}
  {\bibfield  {journal} {\bibinfo  {journal} {Science}\ }\textbf {\bibinfo
  {volume} {305}},\ \bibinfo {pages} {1128} (\bibinfo {year}
  {2004})}\BibitemShut {NoStop}%
\bibitem [{\citenamefont {Randeria}\ \emph {et~al.}(1989)\citenamefont
  {Randeria}, \citenamefont {Duan},\ and\ \citenamefont
  {Shieh}}]{Randeria.1989}%
  \BibitemOpen
  \bibfield  {author} {\bibinfo {author} {\bibfnamefont {M.}~\bibnamefont
  {Randeria}}, \bibinfo {author} {\bibfnamefont {J.-M.}\ \bibnamefont {Duan}},\
  and\ \bibinfo {author} {\bibfnamefont {L.-Y.}\ \bibnamefont {Shieh}},\ }\href
  {https://doi.org/10.1103/PhysRevLett.62.981} {\bibfield  {journal} {\bibinfo
  {journal} {Phys. Rev. Lett.}\ }\textbf {\bibinfo {volume} {62}},\ \bibinfo
  {pages} {981} (\bibinfo {year} {1989})}\BibitemShut {NoStop}%
\bibitem [{\citenamefont {Bloch}\ \emph {et~al.}(2008)\citenamefont {Bloch},
  \citenamefont {Dalibard},\ and\ \citenamefont {Zwerger}}]{Bloch.2008}%
  \BibitemOpen
  \bibfield  {author} {\bibinfo {author} {\bibfnamefont {I.}~\bibnamefont
  {Bloch}}, \bibinfo {author} {\bibfnamefont {J.}~\bibnamefont {Dalibard}},\
  and\ \bibinfo {author} {\bibfnamefont {W.}~\bibnamefont {Zwerger}},\ }\href
  {https://doi.org/10.1103/RevModPhys.80.885} {\bibfield  {journal} {\bibinfo
  {journal} {Rev. Mod. Phys.}\ }\textbf {\bibinfo {volume} {80}},\ \bibinfo
  {pages} {885} (\bibinfo {year} {2008})}\BibitemShut {NoStop}%
\bibitem [{\citenamefont {Giorgini}\ \emph {et~al.}(2008)\citenamefont
  {Giorgini}, \citenamefont {Pitaevskii},\ and\ \citenamefont
  {Stringari}}]{Giorgini.2008}%
  \BibitemOpen
  \bibfield  {author} {\bibinfo {author} {\bibfnamefont {S.}~\bibnamefont
  {Giorgini}}, \bibinfo {author} {\bibfnamefont {L.~P.}\ \bibnamefont
  {Pitaevskii}},\ and\ \bibinfo {author} {\bibfnamefont {S.}~\bibnamefont
  {Stringari}},\ }\href {https://doi.org/10.1103/RevModPhys.80.1215} {\bibfield
   {journal} {\bibinfo  {journal} {Rev. Mod. Phys.}\ }\textbf {\bibinfo
  {volume} {80}},\ \bibinfo {pages} {1215} (\bibinfo {year}
  {2008})}\BibitemShut {NoStop}%
\bibitem [{\citenamefont {Esslinger}(2010)}]{Esslinger.2010}%
  \BibitemOpen
  \bibfield  {author} {\bibinfo {author} {\bibfnamefont {T.}~\bibnamefont
  {Esslinger}},\ }\href
  {https://doi.org/10.1146/annurev-conmatphys-070909-104059} {\bibfield
  {journal} {\bibinfo  {journal} {Annual Review of Condensed Matter Physics}\
  }\textbf {\bibinfo {volume} {1}},\ \bibinfo {pages} {129} (\bibinfo {year}
  {2010})}\BibitemShut {NoStop}%
\bibitem [{\citenamefont {Guan}\ \emph {et~al.}(2013)\citenamefont {Guan},
  \citenamefont {Batchelor},\ and\ \citenamefont {Lee}}]{Guan.2013}%
  \BibitemOpen
  \bibfield  {author} {\bibinfo {author} {\bibfnamefont {X.-W.}\ \bibnamefont
  {Guan}}, \bibinfo {author} {\bibfnamefont {M.~T.}\ \bibnamefont
  {Batchelor}},\ and\ \bibinfo {author} {\bibfnamefont {C.}~\bibnamefont
  {Lee}},\ }\href {https://doi.org/10.1103/RevModPhys.85.1633} {\bibfield
  {journal} {\bibinfo  {journal} {Rev. Mod. Phys.}\ }\textbf {\bibinfo {volume}
  {85}},\ \bibinfo {pages} {1633} (\bibinfo {year} {2013})}\BibitemShut
  {NoStop}%
\bibitem [{\citenamefont {Alford}\ \emph {et~al.}(1998)\citenamefont {Alford},
  \citenamefont {Rajagopal},\ and\ \citenamefont {Wilczek}}]{Alford.1998}%
  \BibitemOpen
  \bibfield  {author} {\bibinfo {author} {\bibfnamefont {M.}~\bibnamefont
  {Alford}}, \bibinfo {author} {\bibfnamefont {K.}~\bibnamefont {Rajagopal}},\
  and\ \bibinfo {author} {\bibfnamefont {F.}~\bibnamefont {Wilczek}},\ }\href
  {https://doi.org/https://doi.org/10.1016/S0370-2693(98)00051-3} {\bibfield
  {journal} {\bibinfo  {journal} {Physics Letters B}\ }\textbf {\bibinfo
  {volume} {422}},\ \bibinfo {pages} {247} (\bibinfo {year}
  {1998})}\BibitemShut {NoStop}%
\bibitem [{\citenamefont {Alford}\ \emph {et~al.}(2008)\citenamefont {Alford},
  \citenamefont {Schmitt}, \citenamefont {Rajagopal},\ and\ \citenamefont
  {Sch\"afer}}]{Alford.2008}%
  \BibitemOpen
  \bibfield  {author} {\bibinfo {author} {\bibfnamefont {M.~G.}\ \bibnamefont
  {Alford}}, \bibinfo {author} {\bibfnamefont {A.}~\bibnamefont {Schmitt}},
  \bibinfo {author} {\bibfnamefont {K.}~\bibnamefont {Rajagopal}},\ and\
  \bibinfo {author} {\bibfnamefont {T.}~\bibnamefont {Sch\"afer}},\ }\href
  {https://doi.org/10.1103/RevModPhys.80.1455} {\bibfield  {journal} {\bibinfo
  {journal} {Rev. Mod. Phys.}\ }\textbf {\bibinfo {volume} {80}},\ \bibinfo
  {pages} {1455} (\bibinfo {year} {2008})}\BibitemShut {NoStop}%
\bibitem [{\citenamefont {Berges}\ and\ \citenamefont
  {Rajagopal}(1999)}]{Berges.1999}%
  \BibitemOpen
  \bibfield  {author} {\bibinfo {author} {\bibfnamefont {J.}~\bibnamefont
  {Berges}}\ and\ \bibinfo {author} {\bibfnamefont {K.}~\bibnamefont
  {Rajagopal}},\ }\href
  {https://doi.org/https://doi.org/10.1016/S0550-3213(98)00620-8} {\bibfield
  {journal} {\bibinfo  {journal} {Nuclear Physics B}\ }\textbf {\bibinfo
  {volume} {538}},\ \bibinfo {pages} {215} (\bibinfo {year}
  {1999})}\BibitemShut {NoStop}%
\bibitem [{\citenamefont {Honerkamp}\ and\ \citenamefont
  {Hofstetter}(2004)}]{Honerkamp.2004}%
  \BibitemOpen
  \bibfield  {author} {\bibinfo {author} {\bibfnamefont {C.}~\bibnamefont
  {Honerkamp}}\ and\ \bibinfo {author} {\bibfnamefont {W.}~\bibnamefont
  {Hofstetter}},\ }\href {https://doi.org/10.1103/PhysRevLett.92.170403}
  {\bibfield  {journal} {\bibinfo  {journal} {Phys. Rev. Lett.}\ }\textbf
  {\bibinfo {volume} {92}},\ \bibinfo {pages} {170403} (\bibinfo {year}
  {2004})}\BibitemShut {NoStop}%
\bibitem [{\citenamefont {He}\ \emph {et~al.}(2006)\citenamefont {He},
  \citenamefont {Jin},\ and\ \citenamefont {Zhuang}}]{He.2006}%
  \BibitemOpen
  \bibfield  {author} {\bibinfo {author} {\bibfnamefont {L.}~\bibnamefont
  {He}}, \bibinfo {author} {\bibfnamefont {M.}~\bibnamefont {Jin}},\ and\
  \bibinfo {author} {\bibfnamefont {P.}~\bibnamefont {Zhuang}},\ }\href
  {https://doi.org/10.1103/PhysRevA.74.033604} {\bibfield  {journal} {\bibinfo
  {journal} {Phys. Rev. A}\ }\textbf {\bibinfo {volume} {74}},\ \bibinfo
  {pages} {033604} (\bibinfo {year} {2006})}\BibitemShut {NoStop}%
\bibitem [{\citenamefont {Rapp}\ \emph {et~al.}(2007)\citenamefont {Rapp},
  \citenamefont {Zar\'and}, \citenamefont {Honerkamp},\ and\ \citenamefont
  {Hofstetter}}]{Rapp.2007}%
  \BibitemOpen
  \bibfield  {author} {\bibinfo {author} {\bibfnamefont {A.}~\bibnamefont
  {Rapp}}, \bibinfo {author} {\bibfnamefont {G.}~\bibnamefont {Zar\'and}},
  \bibinfo {author} {\bibfnamefont {C.}~\bibnamefont {Honerkamp}},\ and\
  \bibinfo {author} {\bibfnamefont {W.}~\bibnamefont {Hofstetter}},\ }\href
  {https://doi.org/10.1103/PhysRevLett.98.160405} {\bibfield  {journal}
  {\bibinfo  {journal} {Phys. Rev. Lett.}\ }\textbf {\bibinfo {volume} {98}},\
  \bibinfo {pages} {160405} (\bibinfo {year} {2007})}\BibitemShut {NoStop}%
\bibitem [{\citenamefont {Gorshkov}\ \emph {et~al.}(2010)\citenamefont
  {Gorshkov}, \citenamefont {Hermele}, \citenamefont {Gurarie}, \citenamefont
  {Xu}, \citenamefont {Julienne}, \citenamefont {Ye}, \citenamefont {Zoller},
  \citenamefont {Demler}, \citenamefont {Lukin},\ and\ \citenamefont
  {Rey}}]{Gorshkov.2010}%
  \BibitemOpen
  \bibfield  {author} {\bibinfo {author} {\bibfnamefont {A.~V.}\ \bibnamefont
  {Gorshkov}}, \bibinfo {author} {\bibfnamefont {M.}~\bibnamefont {Hermele}},
  \bibinfo {author} {\bibfnamefont {V.}~\bibnamefont {Gurarie}}, \bibinfo
  {author} {\bibfnamefont {C.}~\bibnamefont {Xu}}, \bibinfo {author}
  {\bibfnamefont {P.~S.}\ \bibnamefont {Julienne}}, \bibinfo {author}
  {\bibfnamefont {J.}~\bibnamefont {Ye}}, \bibinfo {author} {\bibfnamefont
  {P.}~\bibnamefont {Zoller}}, \bibinfo {author} {\bibfnamefont
  {E.}~\bibnamefont {Demler}}, \bibinfo {author} {\bibfnamefont {M.~D.}\
  \bibnamefont {Lukin}},\ and\ \bibinfo {author} {\bibfnamefont {A.~M.}\
  \bibnamefont {Rey}},\ }\href {https://doi.org/10.1038/nphys1535} {\bibfield
  {journal} {\bibinfo  {journal} {Nature Physics}\ }\textbf {\bibinfo {volume}
  {6}},\ \bibinfo {pages} {289} (\bibinfo {year} {2010})}\BibitemShut {NoStop}%
\bibitem [{\citenamefont {Scazza}\ \emph {et~al.}(2014)\citenamefont {Scazza},
  \citenamefont {Hofrichter}, \citenamefont {H{\"o}fer}, \citenamefont
  {De~Groot}, \citenamefont {Bloch},\ and\ \citenamefont
  {F{\"o}lling}}]{Scazza.2014}%
  \BibitemOpen
  \bibfield  {author} {\bibinfo {author} {\bibfnamefont {F.}~\bibnamefont
  {Scazza}}, \bibinfo {author} {\bibfnamefont {C.}~\bibnamefont {Hofrichter}},
  \bibinfo {author} {\bibfnamefont {M.}~\bibnamefont {H{\"o}fer}}, \bibinfo
  {author} {\bibfnamefont {P.~C.}\ \bibnamefont {De~Groot}}, \bibinfo {author}
  {\bibfnamefont {I.}~\bibnamefont {Bloch}},\ and\ \bibinfo {author}
  {\bibfnamefont {S.}~\bibnamefont {F{\"o}lling}},\ }\href
  {https://doi.org/10.1038/nphys3061} {\bibfield  {journal} {\bibinfo
  {journal} {Nature Physics}\ }\textbf {\bibinfo {volume} {10}},\ \bibinfo
  {pages} {779} (\bibinfo {year} {2014})}\BibitemShut {NoStop}%
\bibitem [{\citenamefont {Cazalilla}\ and\ \citenamefont
  {Rey}(2014)}]{Cazalilla.2014}%
  \BibitemOpen
  \bibfield  {author} {\bibinfo {author} {\bibfnamefont {M.~A.}\ \bibnamefont
  {Cazalilla}}\ and\ \bibinfo {author} {\bibfnamefont {A.~M.}\ \bibnamefont
  {Rey}},\ }\href {https://doi.org/10.1088/0034-4885/77/12/124401} {\bibfield
  {journal} {\bibinfo  {journal} {Reports on Progress in Physics}\ }\textbf
  {\bibinfo {volume} {77}},\ \bibinfo {pages} {124401} (\bibinfo {year}
  {2014})}\BibitemShut {NoStop}%
\bibitem [{\citenamefont {Hofrichter}\ \emph {et~al.}(2016)\citenamefont
  {Hofrichter}, \citenamefont {Riegger}, \citenamefont {Scazza}, \citenamefont
  {H\"ofer}, \citenamefont {Fernandes}, \citenamefont {Bloch},\ and\
  \citenamefont {F\"olling}}]{Hofrichter.2016}%
  \BibitemOpen
  \bibfield  {author} {\bibinfo {author} {\bibfnamefont {C.}~\bibnamefont
  {Hofrichter}}, \bibinfo {author} {\bibfnamefont {L.}~\bibnamefont {Riegger}},
  \bibinfo {author} {\bibfnamefont {F.}~\bibnamefont {Scazza}}, \bibinfo
  {author} {\bibfnamefont {M.}~\bibnamefont {H\"ofer}}, \bibinfo {author}
  {\bibfnamefont {D.~R.}\ \bibnamefont {Fernandes}}, \bibinfo {author}
  {\bibfnamefont {I.}~\bibnamefont {Bloch}},\ and\ \bibinfo {author}
  {\bibfnamefont {S.}~\bibnamefont {F\"olling}},\ }\href
  {https://doi.org/10.1103/PhysRevX.6.021030} {\bibfield  {journal} {\bibinfo
  {journal} {Phys. Rev. X}\ }\textbf {\bibinfo {volume} {6}},\ \bibinfo {pages}
  {021030} (\bibinfo {year} {2016})}\BibitemShut {NoStop}%
\bibitem [{\citenamefont {Huckans}\ \emph {et~al.}(2009)\citenamefont
  {Huckans}, \citenamefont {Williams}, \citenamefont {Hazlett}, \citenamefont
  {Stites},\ and\ \citenamefont {O'Hara}}]{Huckans.2009}%
  \BibitemOpen
  \bibfield  {author} {\bibinfo {author} {\bibfnamefont {J.~H.}\ \bibnamefont
  {Huckans}}, \bibinfo {author} {\bibfnamefont {J.~R.}\ \bibnamefont
  {Williams}}, \bibinfo {author} {\bibfnamefont {E.~L.}\ \bibnamefont
  {Hazlett}}, \bibinfo {author} {\bibfnamefont {R.~W.}\ \bibnamefont
  {Stites}},\ and\ \bibinfo {author} {\bibfnamefont {K.~M.}\ \bibnamefont
  {O'Hara}},\ }\href {https://doi.org/10.1103/PhysRevLett.102.165302}
  {\bibfield  {journal} {\bibinfo  {journal} {Phys. Rev. Lett.}\ }\textbf
  {\bibinfo {volume} {102}},\ \bibinfo {pages} {165302} (\bibinfo {year}
  {2009})}\BibitemShut {NoStop}%
\bibitem [{\citenamefont {Fukuhara}\ \emph {et~al.}(2007)\citenamefont
  {Fukuhara}, \citenamefont {Takasu}, \citenamefont {Kumakura},\ and\
  \citenamefont {Takahashi}}]{Fukuhara.2007}%
  \BibitemOpen
  \bibfield  {author} {\bibinfo {author} {\bibfnamefont {T.}~\bibnamefont
  {Fukuhara}}, \bibinfo {author} {\bibfnamefont {Y.}~\bibnamefont {Takasu}},
  \bibinfo {author} {\bibfnamefont {M.}~\bibnamefont {Kumakura}},\ and\
  \bibinfo {author} {\bibfnamefont {Y.}~\bibnamefont {Takahashi}},\ }\href
  {https://doi.org/10.1103/PhysRevLett.98.030401} {\bibfield  {journal}
  {\bibinfo  {journal} {Phys. Rev. Lett.}\ }\textbf {\bibinfo {volume} {98}},\
  \bibinfo {pages} {030401} (\bibinfo {year} {2007})}\BibitemShut {NoStop}%
\bibitem [{\citenamefont {Taie}\ \emph {et~al.}(2010)\citenamefont {Taie},
  \citenamefont {Takasu}, \citenamefont {Sugawa}, \citenamefont {Yamazaki},
  \citenamefont {Tsujimoto}, \citenamefont {Murakami},\ and\ \citenamefont
  {Takahashi}}]{Taie.2010}%
  \BibitemOpen
  \bibfield  {author} {\bibinfo {author} {\bibfnamefont {S.}~\bibnamefont
  {Taie}}, \bibinfo {author} {\bibfnamefont {Y.}~\bibnamefont {Takasu}},
  \bibinfo {author} {\bibfnamefont {S.}~\bibnamefont {Sugawa}}, \bibinfo
  {author} {\bibfnamefont {R.}~\bibnamefont {Yamazaki}}, \bibinfo {author}
  {\bibfnamefont {T.}~\bibnamefont {Tsujimoto}}, \bibinfo {author}
  {\bibfnamefont {R.}~\bibnamefont {Murakami}},\ and\ \bibinfo {author}
  {\bibfnamefont {Y.}~\bibnamefont {Takahashi}},\ }\href
  {https://doi.org/10.1103/PhysRevLett.105.190401} {\bibfield  {journal}
  {\bibinfo  {journal} {Phys. Rev. Lett.}\ }\textbf {\bibinfo {volume} {105}},\
  \bibinfo {pages} {190401} (\bibinfo {year} {2010})}\BibitemShut {NoStop}%
\bibitem [{\citenamefont {Taie}\ \emph {et~al.}(2012)\citenamefont {Taie},
  \citenamefont {Yamazaki}, \citenamefont {Sugawa},\ and\ \citenamefont
  {Takahashi}}]{Taie.2012}%
  \BibitemOpen
  \bibfield  {author} {\bibinfo {author} {\bibfnamefont {S.}~\bibnamefont
  {Taie}}, \bibinfo {author} {\bibfnamefont {R.}~\bibnamefont {Yamazaki}},
  \bibinfo {author} {\bibfnamefont {S.}~\bibnamefont {Sugawa}},\ and\ \bibinfo
  {author} {\bibfnamefont {Y.}~\bibnamefont {Takahashi}},\ }\href
  {https://doi.org/10.1038/nphys2430} {\bibfield  {journal} {\bibinfo
  {journal} {Nature Physics}\ }\textbf {\bibinfo {volume} {8}},\ \bibinfo
  {pages} {825} (\bibinfo {year} {2012})}\BibitemShut {NoStop}%
\bibitem [{\citenamefont {Zhang}\ \emph {et~al.}(2014)\citenamefont {Zhang},
  \citenamefont {Bishof}, \citenamefont {Bromley}, \citenamefont {Kraus},
  \citenamefont {Safronova}, \citenamefont {Zoller}, \citenamefont {Rey},\ and\
  \citenamefont {Ye}}]{Zhang.2014}%
  \BibitemOpen
  \bibfield  {author} {\bibinfo {author} {\bibfnamefont {X.}~\bibnamefont
  {Zhang}}, \bibinfo {author} {\bibfnamefont {M.}~\bibnamefont {Bishof}},
  \bibinfo {author} {\bibfnamefont {S.~L.}\ \bibnamefont {Bromley}}, \bibinfo
  {author} {\bibfnamefont {C.~V.}\ \bibnamefont {Kraus}}, \bibinfo {author}
  {\bibfnamefont {M.~S.}\ \bibnamefont {Safronova}}, \bibinfo {author}
  {\bibfnamefont {P.}~\bibnamefont {Zoller}}, \bibinfo {author} {\bibfnamefont
  {A.~M.}\ \bibnamefont {Rey}},\ and\ \bibinfo {author} {\bibfnamefont
  {J.}~\bibnamefont {Ye}},\ }\href {https://doi.org/10.1126/science.1254978}
  {\bibfield  {journal} {\bibinfo  {journal} {Science}\ }\textbf {\bibinfo
  {volume} {345}},\ \bibinfo {pages} {1467} (\bibinfo {year}
  {2014})}\BibitemShut {NoStop}%
\bibitem [{\citenamefont {Sonderhouse}\ \emph {et~al.}(2020)\citenamefont
  {Sonderhouse}, \citenamefont {Sanner}, \citenamefont {Hutson}, \citenamefont
  {Goban}, \citenamefont {Bilitewski}, \citenamefont {Yan}, \citenamefont
  {Milner}, \citenamefont {Rey},\ and\ \citenamefont {Ye}}]{Sonderhous.2020}%
  \BibitemOpen
  \bibfield  {author} {\bibinfo {author} {\bibfnamefont {L.}~\bibnamefont
  {Sonderhouse}}, \bibinfo {author} {\bibfnamefont {C.}~\bibnamefont {Sanner}},
  \bibinfo {author} {\bibfnamefont {R.~B.}\ \bibnamefont {Hutson}}, \bibinfo
  {author} {\bibfnamefont {A.}~\bibnamefont {Goban}}, \bibinfo {author}
  {\bibfnamefont {T.}~\bibnamefont {Bilitewski}}, \bibinfo {author}
  {\bibfnamefont {L.}~\bibnamefont {Yan}}, \bibinfo {author} {\bibfnamefont
  {W.~R.}\ \bibnamefont {Milner}}, \bibinfo {author} {\bibfnamefont {A.~M.}\
  \bibnamefont {Rey}},\ and\ \bibinfo {author} {\bibfnamefont {J.}~\bibnamefont
  {Ye}},\ }\href {https://doi.org/10.1038/s41567-020-0986-6} {\bibfield
  {journal} {\bibinfo  {journal} {Nature Physics}\ }\textbf {\bibinfo {volume}
  {16}},\ \bibinfo {pages} {1216} (\bibinfo {year} {2020})}\BibitemShut
  {NoStop}%
\bibitem [{\citenamefont {Ulbricht}\ \emph {et~al.}(2010)\citenamefont
  {Ulbricht}, \citenamefont {Molina}, \citenamefont {Thomale},\ and\
  \citenamefont {Schmitteckert}}]{Ulbricht.2010}%
  \BibitemOpen
  \bibfield  {author} {\bibinfo {author} {\bibfnamefont {T.}~\bibnamefont
  {Ulbricht}}, \bibinfo {author} {\bibfnamefont {R.~A.}\ \bibnamefont
  {Molina}}, \bibinfo {author} {\bibfnamefont {R.}~\bibnamefont {Thomale}},\
  and\ \bibinfo {author} {\bibfnamefont {P.}~\bibnamefont {Schmitteckert}},\
  }\href {https://doi.org/10.1103/PhysRevA.82.011603} {\bibfield  {journal}
  {\bibinfo  {journal} {Phys. Rev. A}\ }\textbf {\bibinfo {volume} {82}},\
  \bibinfo {pages} {011603} (\bibinfo {year} {2010})}\BibitemShut {NoStop}%
\bibitem [{\citenamefont {Dutta}\ \emph {et~al.}(2015)\citenamefont {Dutta},
  \citenamefont {Gajda}, \citenamefont {Hauke}, \citenamefont {Lewenstein},
  \citenamefont {Lühmann}, \citenamefont {Malomed}, \citenamefont
  {Sowi{\'{n}}ski},\ and\ \citenamefont {Zakrzewski}}]{Dutta.2015}%
  \BibitemOpen
  \bibfield  {author} {\bibinfo {author} {\bibfnamefont {O.}~\bibnamefont
  {Dutta}}, \bibinfo {author} {\bibfnamefont {M.}~\bibnamefont {Gajda}},
  \bibinfo {author} {\bibfnamefont {P.}~\bibnamefont {Hauke}}, \bibinfo
  {author} {\bibfnamefont {M.}~\bibnamefont {Lewenstein}}, \bibinfo {author}
  {\bibfnamefont {D.-S.}\ \bibnamefont {Lühmann}}, \bibinfo {author}
  {\bibfnamefont {B.~A.}\ \bibnamefont {Malomed}}, \bibinfo {author}
  {\bibfnamefont {T.}~\bibnamefont {Sowi{\'{n}}ski}},\ and\ \bibinfo {author}
  {\bibfnamefont {J.}~\bibnamefont {Zakrzewski}},\ }\href
  {https://doi.org/10.1088/0034-4885/78/6/066001} {\bibfield  {journal}
  {\bibinfo  {journal} {Reports on Progress in Physics}\ }\textbf {\bibinfo
  {volume} {78}},\ \bibinfo {pages} {066001} (\bibinfo {year}
  {2015})}\BibitemShut {NoStop}%
\bibitem [{\citenamefont {Winkler}\ \emph {et~al.}(2006)\citenamefont
  {Winkler}, \citenamefont {Thalhammer}, \citenamefont {Lang}, \citenamefont
  {Grimm}, \citenamefont {Hecker~Denschlag}, \citenamefont {Daley},
  \citenamefont {Kantian}, \citenamefont {B{\"u}chler},\ and\ \citenamefont
  {Zoller}}]{Winkler.2006}%
  \BibitemOpen
  \bibfield  {author} {\bibinfo {author} {\bibfnamefont {K.}~\bibnamefont
  {Winkler}}, \bibinfo {author} {\bibfnamefont {G.}~\bibnamefont {Thalhammer}},
  \bibinfo {author} {\bibfnamefont {F.}~\bibnamefont {Lang}}, \bibinfo {author}
  {\bibfnamefont {R.}~\bibnamefont {Grimm}}, \bibinfo {author} {\bibfnamefont
  {J.}~\bibnamefont {Hecker~Denschlag}}, \bibinfo {author} {\bibfnamefont
  {A.~J.}\ \bibnamefont {Daley}}, \bibinfo {author} {\bibfnamefont
  {A.}~\bibnamefont {Kantian}}, \bibinfo {author} {\bibfnamefont {H.~P.}\
  \bibnamefont {B{\"u}chler}},\ and\ \bibinfo {author} {\bibfnamefont
  {P.}~\bibnamefont {Zoller}},\ }\href {https://doi.org/10.1038/nature04918}
  {\bibfield  {journal} {\bibinfo  {journal} {Nature}\ }\textbf {\bibinfo
  {volume} {441}},\ \bibinfo {pages} {853} (\bibinfo {year}
  {2006})}\BibitemShut {NoStop}%
\bibitem [{\citenamefont {Wang}\ and\ \citenamefont {Liang}(2010)}]{Wang.2010}%
  \BibitemOpen
  \bibfield  {author} {\bibinfo {author} {\bibfnamefont {Y.-M.}\ \bibnamefont
  {Wang}}\ and\ \bibinfo {author} {\bibfnamefont {J.-Q.}\ \bibnamefont
  {Liang}},\ }\href {https://doi.org/10.1103/PhysRevA.81.045601} {\bibfield
  {journal} {\bibinfo  {journal} {Phys. Rev. A}\ }\textbf {\bibinfo {volume}
  {81}},\ \bibinfo {pages} {045601} (\bibinfo {year} {2010})}\BibitemShut
  {NoStop}%
\bibitem [{\citenamefont {Trotzky}\ \emph {et~al.}(2012)\citenamefont
  {Trotzky}, \citenamefont {Chen}, \citenamefont {Flesch}, \citenamefont
  {McCulloch}, \citenamefont {Schollw{\"o}ck}, \citenamefont {Eisert},\ and\
  \citenamefont {Bloch}}]{Trotzky.2012}%
  \BibitemOpen
  \bibfield  {author} {\bibinfo {author} {\bibfnamefont {S.}~\bibnamefont
  {Trotzky}}, \bibinfo {author} {\bibfnamefont {Y.-A.}\ \bibnamefont {Chen}},
  \bibinfo {author} {\bibfnamefont {A.}~\bibnamefont {Flesch}}, \bibinfo
  {author} {\bibfnamefont {I.~P.}\ \bibnamefont {McCulloch}}, \bibinfo {author}
  {\bibfnamefont {U.}~\bibnamefont {Schollw{\"o}ck}}, \bibinfo {author}
  {\bibfnamefont {J.}~\bibnamefont {Eisert}},\ and\ \bibinfo {author}
  {\bibfnamefont {I.}~\bibnamefont {Bloch}},\ }\href
  {https://doi.org/10.1038/nphys2232} {\bibfield  {journal} {\bibinfo
  {journal} {Nature Physics}\ }\textbf {\bibinfo {volume} {8}},\ \bibinfo
  {pages} {325} (\bibinfo {year} {2012})}\BibitemShut {NoStop}%
\bibitem [{\citenamefont {Vidal}(2004)}]{Vidal.2004}%
  \BibitemOpen
  \bibfield  {author} {\bibinfo {author} {\bibfnamefont {G.}~\bibnamefont
  {Vidal}},\ }\href {https://doi.org/10.1103/PhysRevLett.93.040502} {\bibfield
  {journal} {\bibinfo  {journal} {Phys. Rev. Lett.}\ }\textbf {\bibinfo
  {volume} {93}},\ \bibinfo {pages} {040502} (\bibinfo {year}
  {2004})}\BibitemShut {NoStop}%
\bibitem [{\citenamefont {Vidal}(2007)}]{Vidal.2007}%
  \BibitemOpen
  \bibfield  {author} {\bibinfo {author} {\bibfnamefont {G.}~\bibnamefont
  {Vidal}},\ }\href {https://doi.org/10.1103/PhysRevLett.98.070201} {\bibfield
  {journal} {\bibinfo  {journal} {Phys. Rev. Lett.}\ }\textbf {\bibinfo
  {volume} {98}},\ \bibinfo {pages} {070201} (\bibinfo {year}
  {2007})}\BibitemShut {NoStop}%
\bibitem [{\citenamefont {Werner}\ \emph {et~al.}(2020)\citenamefont {Werner},
  \citenamefont {Moca}, \citenamefont {Legeza},\ and\ \citenamefont
  {Zar\'and}}]{Werner.2020}%
  \BibitemOpen
  \bibfield  {author} {\bibinfo {author} {\bibfnamefont {M.~A.}\ \bibnamefont
  {Werner}}, \bibinfo {author} {\bibfnamefont {C.~P.}\ \bibnamefont {Moca}},
  \bibinfo {author} {\bibfnamefont {O.}~\bibnamefont {Legeza}},\ and\ \bibinfo
  {author} {\bibfnamefont {G.}~\bibnamefont {Zar\'and}},\ }\href
  {https://doi.org/10.1103/PhysRevB.102.155108} {\bibfield  {journal} {\bibinfo
   {journal} {Phys. Rev. B}\ }\textbf {\bibinfo {volume} {102}},\ \bibinfo
  {pages} {155108} (\bibinfo {year} {2020})}\BibitemShut {NoStop}%
\bibitem [{\citenamefont {Takahashi}(1977)}]{Takahashi.1977}%
  \BibitemOpen
  \bibfield  {author} {\bibinfo {author} {\bibfnamefont {M.}~\bibnamefont
  {Takahashi}},\ }\href {https://doi.org/10.1088/0022-3719/10/8/031} {\bibfield
   {journal} {\bibinfo  {journal} {Journal of Physics C: Solid State Physics}\
  }\textbf {\bibinfo {volume} {10}},\ \bibinfo {pages} {1289} (\bibinfo {year}
  {1977})}\BibitemShut {NoStop}%
\bibitem [{\citenamefont {Valmispild}\ \emph {et~al.}(2020)\citenamefont
  {Valmispild}, \citenamefont {Dutreix}, \citenamefont {Eckstein},
  \citenamefont {Katsnelson}, \citenamefont {Lichtenstein},\ and\ \citenamefont
  {Stepanov}}]{Valmispild.2020}%
  \BibitemOpen
  \bibfield  {author} {\bibinfo {author} {\bibfnamefont {V.~N.}\ \bibnamefont
  {Valmispild}}, \bibinfo {author} {\bibfnamefont {C.}~\bibnamefont {Dutreix}},
  \bibinfo {author} {\bibfnamefont {M.}~\bibnamefont {Eckstein}}, \bibinfo
  {author} {\bibfnamefont {M.~I.}\ \bibnamefont {Katsnelson}}, \bibinfo
  {author} {\bibfnamefont {A.~I.}\ \bibnamefont {Lichtenstein}},\ and\ \bibinfo
  {author} {\bibfnamefont {E.~A.}\ \bibnamefont {Stepanov}},\ }\href
  {https://doi.org/10.1103/PhysRevB.102.220301} {\bibfield  {journal} {\bibinfo
   {journal} {Phys. Rev. B}\ }\textbf {\bibinfo {volume} {102}},\ \bibinfo
  {pages} {220301} (\bibinfo {year} {2020})}\BibitemShut {NoStop}%
\bibitem [{\citenamefont {Strohmaier}\ \emph {et~al.}(2010)\citenamefont
  {Strohmaier}, \citenamefont {Greif}, \citenamefont {J\"ordens}, \citenamefont
  {Tarruell}, \citenamefont {Moritz}, \citenamefont {Esslinger}, \citenamefont
  {Sensarma}, \citenamefont {Pekker}, \citenamefont {Altman},\ and\
  \citenamefont {Demler}}]{Strohmaier.2010}%
  \BibitemOpen
  \bibfield  {author} {\bibinfo {author} {\bibfnamefont {N.}~\bibnamefont
  {Strohmaier}}, \bibinfo {author} {\bibfnamefont {D.}~\bibnamefont {Greif}},
  \bibinfo {author} {\bibfnamefont {R.}~\bibnamefont {J\"ordens}}, \bibinfo
  {author} {\bibfnamefont {L.}~\bibnamefont {Tarruell}}, \bibinfo {author}
  {\bibfnamefont {H.}~\bibnamefont {Moritz}}, \bibinfo {author} {\bibfnamefont
  {T.}~\bibnamefont {Esslinger}}, \bibinfo {author} {\bibfnamefont
  {R.}~\bibnamefont {Sensarma}}, \bibinfo {author} {\bibfnamefont
  {D.}~\bibnamefont {Pekker}}, \bibinfo {author} {\bibfnamefont
  {E.}~\bibnamefont {Altman}},\ and\ \bibinfo {author} {\bibfnamefont
  {E.}~\bibnamefont {Demler}},\ }\href
  {https://doi.org/10.1103/PhysRevLett.104.080401} {\bibfield  {journal}
  {\bibinfo  {journal} {Phys. Rev. Lett.}\ }\textbf {\bibinfo {volume} {104}},\
  \bibinfo {pages} {080401} (\bibinfo {year} {2010})}\BibitemShut {NoStop}%
\bibitem [{\citenamefont {Rapp}\ \emph {et~al.}(2010)\citenamefont {Rapp},
  \citenamefont {Mandt},\ and\ \citenamefont {Rosch}}]{Rapp.2010}%
  \BibitemOpen
  \bibfield  {author} {\bibinfo {author} {\bibfnamefont {A.}~\bibnamefont
  {Rapp}}, \bibinfo {author} {\bibfnamefont {S.}~\bibnamefont {Mandt}},\ and\
  \bibinfo {author} {\bibfnamefont {A.}~\bibnamefont {Rosch}},\ }\href
  {https://doi.org/10.1103/PhysRevLett.105.220405} {\bibfield  {journal}
  {\bibinfo  {journal} {Phys. Rev. Lett.}\ }\textbf {\bibinfo {volume} {105}},\
  \bibinfo {pages} {220405} (\bibinfo {year} {2010})}\BibitemShut {NoStop}%
\bibitem [{\citenamefont {Rapp}(2013)}]{Rapp.2013}%
  \BibitemOpen
  \bibfield  {author} {\bibinfo {author} {\bibfnamefont {A.}~\bibnamefont
  {Rapp}},\ }\href {https://doi.org/10.1103/PhysRevA.87.043611} {\bibfield
  {journal} {\bibinfo  {journal} {Phys. Rev. A}\ }\textbf {\bibinfo {volume}
  {87}},\ \bibinfo {pages} {043611} (\bibinfo {year} {2013})}\BibitemShut
  {NoStop}%
\bibitem [{\citenamefont {Deutsch}(2018)}]{Deutsch.2018}%
  \BibitemOpen
  \bibfield  {author} {\bibinfo {author} {\bibfnamefont {J.~M.}\ \bibnamefont
  {Deutsch}},\ }\href {https://doi.org/10.1088/1361-6633/aac9f1} {\bibfield
  {journal} {\bibinfo  {journal} {Reports on Progress in Physics}\ }\textbf
  {\bibinfo {volume} {81}},\ \bibinfo {pages} {082001} (\bibinfo {year}
  {2018})}\BibitemShut {NoStop}%
\bibitem [{\citenamefont {Buchta}\ \emph {et~al.}(2007)\citenamefont {Buchta},
  \citenamefont {Legeza}, \citenamefont {Szirmai},\ and\ \citenamefont
  {S{\'o}lyom}}]{Buchta.2007}%
  \BibitemOpen
  \bibfield  {author} {\bibinfo {author} {\bibfnamefont {K.}~\bibnamefont
  {Buchta}}, \bibinfo {author} {\bibfnamefont {{\"O}.}~\bibnamefont {Legeza}},
  \bibinfo {author} {\bibfnamefont {E.}~\bibnamefont {Szirmai}},\ and\ \bibinfo
  {author} {\bibfnamefont {J.}~\bibnamefont {S{\'o}lyom}},\ }\href
  {https://doi.org/10.1103/PhysRevB.75.155108} {\bibfield  {journal} {\bibinfo
  {journal} {Phys. Rev. B}\ }\textbf {\bibinfo {volume} {75}},\ \bibinfo
  {pages} {155108} (\bibinfo {year} {2007})}\BibitemShut {NoStop}%
\bibitem [{\citenamefont {Corboz}\ \emph {et~al.}(2012)\citenamefont {Corboz},
  \citenamefont {Lajk{\'o}}, \citenamefont {L{\"a}uchli}, \citenamefont
  {Penc},\ and\ \citenamefont {Mila}}]{Corboz.2012}%
  \BibitemOpen
  \bibfield  {author} {\bibinfo {author} {\bibfnamefont {P.}~\bibnamefont
  {Corboz}}, \bibinfo {author} {\bibfnamefont {M.}~\bibnamefont {Lajk{\'o}}},
  \bibinfo {author} {\bibfnamefont {A.~M.}\ \bibnamefont {L{\"a}uchli}},
  \bibinfo {author} {\bibfnamefont {K.}~\bibnamefont {Penc}},\ and\ \bibinfo
  {author} {\bibfnamefont {F.}~\bibnamefont {Mila}},\ }\href
  {https://doi.org/10.1103/PhysRevX.2.041013} {\bibfield  {journal} {\bibinfo
  {journal} {Phys. Rev. X}\ }\textbf {\bibinfo {volume} {2}},\ \bibinfo {pages}
  {041013} (\bibinfo {year} {2012})}\BibitemShut {NoStop}%
\bibitem [{\citenamefont {Calabrese}\ and\ \citenamefont
  {Cardy}(2009)}]{Calabrese.2009}%
  \BibitemOpen
  \bibfield  {author} {\bibinfo {author} {\bibfnamefont {P.}~\bibnamefont
  {Calabrese}}\ and\ \bibinfo {author} {\bibfnamefont {J.}~\bibnamefont
  {Cardy}},\ }\href {https://doi.org/10.1088/1751-8113/42/50/504005} {\bibfield
   {journal} {\bibinfo  {journal} {Journal of Physics A: Mathematical and
  Theoretical}\ }\textbf {\bibinfo {volume} {42}},\ \bibinfo {pages} {504005}
  (\bibinfo {year} {2009})}\BibitemShut {NoStop}%
\bibitem [{\citenamefont {Zhou}\ and\ \citenamefont
  {Nahum}(2020)}]{Nahum.2020}%
  \BibitemOpen
  \bibfield  {author} {\bibinfo {author} {\bibfnamefont {T.}~\bibnamefont
  {Zhou}}\ and\ \bibinfo {author} {\bibfnamefont {A.}~\bibnamefont {Nahum}},\
  }\href {https://doi.org/10.1103/PhysRevX.10.031066} {\bibfield  {journal}
  {\bibinfo  {journal} {Phys. Rev. X}\ }\textbf {\bibinfo {volume} {10}},\
  \bibinfo {pages} {031066} (\bibinfo {year} {2020})}\BibitemShut {NoStop}%
\bibitem [{\citenamefont {Altman}\ \emph {et~al.}(2004)\citenamefont {Altman},
  \citenamefont {Demler},\ and\ \citenamefont {Lukin}}]{Altman.2004}%
  \BibitemOpen
  \bibfield  {author} {\bibinfo {author} {\bibfnamefont {E.}~\bibnamefont
  {Altman}}, \bibinfo {author} {\bibfnamefont {E.}~\bibnamefont {Demler}},\
  and\ \bibinfo {author} {\bibfnamefont {M.~D.}\ \bibnamefont {Lukin}},\ }\href
  {https://doi.org/10.1103/PhysRevA.70.013603} {\bibfield  {journal} {\bibinfo
  {journal} {Phys. Rev. A}\ }\textbf {\bibinfo {volume} {70}},\ \bibinfo
  {pages} {013603} (\bibinfo {year} {2004})}\BibitemShut {NoStop}%
\bibitem [{\citenamefont {Gopalakrishnan}\ \emph {et~al.}(2018)\citenamefont
  {Gopalakrishnan}, \citenamefont {Huse}, \citenamefont {Khemani},\ and\
  \citenamefont {Vasseur}}]{Gopalakrishna.2018}%
  \BibitemOpen
  \bibfield  {author} {\bibinfo {author} {\bibfnamefont {S.}~\bibnamefont
  {Gopalakrishnan}}, \bibinfo {author} {\bibfnamefont {D.~A.}\ \bibnamefont
  {Huse}}, \bibinfo {author} {\bibfnamefont {V.}~\bibnamefont {Khemani}},\ and\
  \bibinfo {author} {\bibfnamefont {R.}~\bibnamefont {Vasseur}},\ }\href
  {https://doi.org/10.1103/PhysRevB.98.220303} {\bibfield  {journal} {\bibinfo
  {journal} {Phys. Rev. B}\ }\textbf {\bibinfo {volume} {98}},\ \bibinfo
  {pages} {220303} (\bibinfo {year} {2018})}\BibitemShut {NoStop}%
\bibitem [{\citenamefont {Adams}\ \emph {et~al.}(2012)\citenamefont {Adams},
  \citenamefont {Carr}, \citenamefont {Schäfer}, \citenamefont {Steinberg},\
  and\ \citenamefont {Thomas}}]{Adams.2012}%
  \BibitemOpen
  \bibfield  {author} {\bibinfo {author} {\bibfnamefont {A.}~\bibnamefont
  {Adams}}, \bibinfo {author} {\bibfnamefont {L.~D.}\ \bibnamefont {Carr}},
  \bibinfo {author} {\bibfnamefont {T.}~\bibnamefont {Schäfer}}, \bibinfo
  {author} {\bibfnamefont {P.}~\bibnamefont {Steinberg}},\ and\ \bibinfo
  {author} {\bibfnamefont {J.~E.}\ \bibnamefont {Thomas}},\ }\href
  {https://doi.org/10.1088/1367-2630/14/11/115009} {\bibfield  {journal}
  {\bibinfo  {journal} {New Journal of Physics}\ }\textbf {\bibinfo {volume}
  {14}},\ \bibinfo {pages} {115009} (\bibinfo {year} {2012})}\BibitemShut
  {NoStop}%
\bibitem [{\citenamefont {Daley}\ \emph {et~al.}(2009)\citenamefont {Daley},
  \citenamefont {Taylor}, \citenamefont {Diehl}, \citenamefont {Baranov},\ and\
  \citenamefont {Zoller}}]{Daley.2009}%
  \BibitemOpen
  \bibfield  {author} {\bibinfo {author} {\bibfnamefont {A.~J.}\ \bibnamefont
  {Daley}}, \bibinfo {author} {\bibfnamefont {J.~M.}\ \bibnamefont {Taylor}},
  \bibinfo {author} {\bibfnamefont {S.}~\bibnamefont {Diehl}}, \bibinfo
  {author} {\bibfnamefont {M.}~\bibnamefont {Baranov}},\ and\ \bibinfo {author}
  {\bibfnamefont {P.}~\bibnamefont {Zoller}},\ }\href
  {https://doi.org/10.1103/PhysRevLett.102.040402} {\bibfield  {journal}
  {\bibinfo  {journal} {Phys. Rev. Lett.}\ }\textbf {\bibinfo {volume} {102}},\
  \bibinfo {pages} {040402} (\bibinfo {year} {2009})}\BibitemShut {NoStop}%
\bibitem [{\citenamefont {Chen}\ \emph {et~al.}(2022)\citenamefont {Chen},
  \citenamefont {Duda}, \citenamefont {Schindewolf}, \citenamefont {Bause},
  \citenamefont {Bloch},\ and\ \citenamefont {Luo}}]{Chen.2022}%
  \BibitemOpen
  \bibfield  {author} {\bibinfo {author} {\bibfnamefont {X.-Y.}\ \bibnamefont
  {Chen}}, \bibinfo {author} {\bibfnamefont {M.}~\bibnamefont {Duda}}, \bibinfo
  {author} {\bibfnamefont {A.}~\bibnamefont {Schindewolf}}, \bibinfo {author}
  {\bibfnamefont {R.}~\bibnamefont {Bause}}, \bibinfo {author} {\bibfnamefont
  {I.}~\bibnamefont {Bloch}},\ and\ \bibinfo {author} {\bibfnamefont {X.-Y.}\
  \bibnamefont {Luo}},\ }\href {https://doi.org/10.1103/PhysRevLett.128.153401}
  {\bibfield  {journal} {\bibinfo  {journal} {Phys. Rev. Lett.}\ }\textbf
  {\bibinfo {volume} {128}},\ \bibinfo {pages} {153401} (\bibinfo {year}
  {2022})}\BibitemShut {NoStop}%
\bibitem [{\citenamefont {Cheong}\ and\ \citenamefont
  {Henley}(2009)}]{Cheong.2009}%
  \BibitemOpen
  \bibfield  {author} {\bibinfo {author} {\bibfnamefont {S.-A.}\ \bibnamefont
  {Cheong}}\ and\ \bibinfo {author} {\bibfnamefont {C.~L.}\ \bibnamefont
  {Henley}},\ }\href {https://doi.org/10.1103/PhysRevB.80.165124} {\bibfield
  {journal} {\bibinfo  {journal} {Phys. Rev. B}\ }\textbf {\bibinfo {volume}
  {80}},\ \bibinfo {pages} {165124} (\bibinfo {year} {2009})}\BibitemShut
  {NoStop}%
\bibitem [{\citenamefont {MacDonald}\ \emph {et~al.}(1988)\citenamefont
  {MacDonald}, \citenamefont {Girvin},\ and\ \citenamefont
  {Yoshioka}}]{MacDonald.1988}%
  \BibitemOpen
  \bibfield  {author} {\bibinfo {author} {\bibfnamefont {A.~H.}\ \bibnamefont
  {MacDonald}}, \bibinfo {author} {\bibfnamefont {S.~M.}\ \bibnamefont
  {Girvin}},\ and\ \bibinfo {author} {\bibfnamefont {D.}~\bibnamefont
  {Yoshioka}},\ }\href {https://doi.org/10.1103/PhysRevB.37.9753} {\bibfield
  {journal} {\bibinfo  {journal} {Phys. Rev. B}\ }\textbf {\bibinfo {volume}
  {37}},\ \bibinfo {pages} {9753} (\bibinfo {year} {1988})}\BibitemShut
  {NoStop}%
\bibitem [{\citenamefont {Cai}\ \emph {et~al.}(2021)\citenamefont {Cai},
  \citenamefont {Yang}, \citenamefont {Shi}, \citenamefont {Lee}, \citenamefont
  {Andrei},\ and\ \citenamefont {Guan}}]{Cai.2021}%
  \BibitemOpen
  \bibfield  {author} {\bibinfo {author} {\bibfnamefont {X.}~\bibnamefont
  {Cai}}, \bibinfo {author} {\bibfnamefont {H.}~\bibnamefont {Yang}}, \bibinfo
  {author} {\bibfnamefont {H.-L.}\ \bibnamefont {Shi}}, \bibinfo {author}
  {\bibfnamefont {C.}~\bibnamefont {Lee}}, \bibinfo {author} {\bibfnamefont
  {N.}~\bibnamefont {Andrei}},\ and\ \bibinfo {author} {\bibfnamefont {X.-W.}\
  \bibnamefont {Guan}},\ }\href
  {https://doi.org/10.1103/PhysRevLett.127.100406} {\bibfield  {journal}
  {\bibinfo  {journal} {Phys. Rev. Lett.}\ }\textbf {\bibinfo {volume} {127}},\
  \bibinfo {pages} {100406} (\bibinfo {year} {2021})}\BibitemShut {NoStop}%
\end{thebibliography}%


\null
\vskip0.2cm
{\parindent=0pt
\textbf{\large Acknowledgements}
}

This research is supported by the National Research,
Development and Innovation Office - NKFIH through research grants Nos.  K134983,  K138606 and SNN139581,
within the Quantum National Laboratory of Hungary program (Project No. 2017-1.2.1-NKP-2017-00001).
M.A.W has also been supported by the \' UNKP-21-4-II New National Excellence Program of the National Research,
Development and Innovation Office - NKFIH. C.P.M  acknowledges
support by the Ministry of Research, Innovation and Digitization, CNCS/CCCDI–UEFISCDI, under
projects number PN-III-P4-ID-PCE-2020-0277 and the project for funding
the excellence, contract No. 29 PFE/30.12.2021.
O.L. acknowledges support from the Hans Fischer Senior Fellowship programme
funded by the Technical University of Munich -- Institute for Advanced Study and
from the Center for Scalable and Predictive methods for Excitation and Correlated phenomena (SPEC), funded as part of the Computational Chemical Sciences Program by the U.S.~Department of Energy (DOE), Office of Science, Office of Basic Energy Sciences, Division of Chemical Sciences, Geosciences, and Biosciences at Pacific Northwest National Laboratory.

\null
\vskip0.2cm
{\parindent=0pt
\textbf{\large Author contributions}
\\
M.A.W., C.P.M., Ö.L.,  and G.Z. developed the non-Abelian MPS software. M.A.W. performed the numerical simulations, G.Z. 
conceived and coordinated the project. M.A.W.,  C.P.M.,  G.Z., and  M.K., carried out the analytical calculations. 
G.Z. and B.D. drafted the manuscript, and all  authors contributed to its final version, and participated in 
the interpretation of the results.}

\null
\vskip0.2cm
{\parindent=0pt
\textbf{\large Competing interest}}
\\
The authors declare no competing interests.

\end{document}